%% file: SMP-16-005_temp.tex
\pdfoutput=1

\documentclass[11pt,twoside,a4paper,cmspaper,final,collab]{cms-tdr}
\def\svnVersion{437279P}\def\cmsCernNoTag{CERN-EP-2017-142}\def\cmsCernDate{\today}\def\cmsMessage{Published in Physical Review D as \href{http://dx.doi.org/10.1103/PhysRevD.96.072005}{\doi{10.1103/PhysRevD.96.072005.}}}

\begin{document}\cmsNoteHeader{SMP-16-005}

\hyphenation{had-ron-i-za-tion}
\hyphenation{cal-or-i-me-ter}
\hyphenation{de-vices}
\RCS$Revision: 428943 $
\RCS$HeadURL: svn+ssh://svn.cern.ch/reps/tdr2/papers/SMP-16-005/trunk/SMP-16-005.tex $
\RCS$Id: SMP-16-005.tex 428943 2017-10-11 14:55:56Z ocalan $
\newlength\cmsFigWidth
\ifthenelse{\boolean{cms@external}}{\setlength\cmsFigWidth{0.85\columnwidth}}{\setlength\cmsFigWidth{0.4\textwidth}}
\ifthenelse{\boolean{cms@external}}{\providecommand{\cmsLeft}{top\xspace}}{\providecommand{\cmsLeft}{left\xspace}}
\ifthenelse{\boolean{cms@external}}{\providecommand{\cmsRight}{bottom\xspace}}{\providecommand{\cmsRight}{right\xspace}}
\ifthenelse{\boolean{cms@italic}}{\providecommand{\ghmW}{\ensuremath{W}\xspace}}{\providecommand{\ghmW}{\ensuremath{\mathrm{W}}\xspace}}
\providecommand{\PV}{\ensuremath{\cmsSymbolFace{V}}\xspace}
\newcommand{\x}{\ensuremath{\phantom{0}}}

\cmsNoteHeader{SMP-16-005}
\title{Measurement of the differential cross sections for the associated production of a \texorpdfstring{\ghmW~boson}{W~boson} and jets in proton-proton collisions at \texorpdfstring{$\sqrt{s} = 13\TeV$}{sqrt(s) = 13 TeV}}

\date{\today}

\abstract{A measurement of the differential cross sections for a \PW boson produced in association with jets in the muon decay channel is presented. The measurement is based on 13\TeV proton-proton collision data corresponding to an integrated luminosity of 2.2\fbinv, recorded by the CMS detector at the LHC. The cross sections are reported as functions of jet multiplicity, jet transverse momentum $\pt$, jet rapidity, the scalar $\pt$ sum of the jets, and angular correlations between the muon and each jet for different jet multiplicities. The measured cross sections are in agreement with predictions that include multileg leading-order (LO) and next-to-LO matrix element calculations interfaced with parton showers, as well as a next-to-next-to-LO calculation for the \PW boson and one jet production.
}
\hypersetup{%
pdfauthor={CMS Collaboration},%
pdftitle={Measurement of the differential cross sections for the associated production of a W boson and jets in proton-proton collisions at sqrt(s) = 13 TeV},%
pdfsubject={CMS},%
pdfkeywords={CMS, physics, software, computing}}

\maketitle

\section{Introduction}
\label{introduction}
The high center-of-mass energy of collisions at the CERN LHC facilitates the production of events with an electroweak boson in association with a high transverse momentum $\pt$ jet or high-multiplicity multijet final state. Measurements of vector boson production in association with jets provide important tests of perturbative quantum
chromodynamics (QCD). Such measurements lead to a better understanding of the strong interaction and of the proton structure. Furthermore, events containing a massive vector boson and jets are important backgrounds for a number of standard model (SM) processes (single top quark, $\ttbar$, vector boson fusion, $\PW\PW$ scattering, and Higgs boson production), as well as for physics beyond the SM, such as supersymmetry. Leptonic decay modes of the vector boson are often used in the measurements of SM processes and searches for physics beyond the SM because they provide sufficient signal data with clean signatures and a low level of background.

The differential cross sections for the production of a $\PW$ boson in association with jets ($\PW$+jets) have been measured by the CMS Collaboration using data collected at center-of-mass energies of 7\TeV~\cite{Khachatryan:2014uva} and 8\TeV~\cite{Khachatryan:2016fue} and by the ATLAS Collaboration at 7\TeV~\cite{Aad:2014qxa} at the LHC. In this paper we present the first differential cross section measurement of the $\PW$+jets process at a center-of-mass energy of 13\TeV. The muon decay channel ($\PW(\PGm\PGn)$+jets) is used; the corresponding decay channel of the $\PW$ boson into an electron and a neutrino is not used in this analysis because a higher momentum threshold was applied to the electron when acquiring data. This measurement is based on 13\TeV proton-proton (pp) collision data recorded by the CMS experiment during 2015 and corresponding to an integrated luminosity of 2.2\fbinv. The differential cross sections are measured as functions of the exclusive ($\PW$+$N$-jets) and inclusive ($\PW$+${\geq}N$-jets) jet multiplicities up to $N=6$. The differential cross sections are also measured up to multiplicities of four inclusive jets as functions of the jet $\pt$, the jet rapidity $|y|$, the scalar $\pt$ sum of the jets $\HT$, and of the azimuthal correlations between the muon and the $i$th jet from the $\pt$-ordered list of jets in the event $\Delta\phi(\text{$\PGm$, $j_{i}$})$.

In addition, the differential cross section is measured as a function of the angular distance between the muon and the closest jet $\Delta R(\text{$\PGm$, closest jet})$ for events with one or more jets, where $\Delta R = \sqrt{\smash[b]{(\Delta\eta)^2 + (\Delta\phi)^2}}$ and $\Delta\eta$ is the pseudorapidity $\eta$ separation. The $\Delta R(\text{$\PGm$, closest jet})$ variable separates the process of the electroweak emission of real $\PW$ bosons from an initial- or final-state quark, which was recently studied by the ATLAS Collaboration with 8\TeV data~\cite{atlasrealw}. The contribution of electroweak radiative processes to the measurement of $\PW$+jets becomes significant with the increasing center-of-mass energy of collisions, leading to an enhancement in the collinear region of the distribution of the angular distance between the $\PW$ boson and the closest jet~\cite{Baur:2006sn,Christiansen:2014kba,Krauss:2014yaa,Boughezal:2015oga,Boughezal:2016dtm,Christiansen:2015jpa}.

The measured cross sections are compared with the predictions from Monte Carlo (MC) event generators that use a leading-order (LO), or a next-to-LO (NLO) matrix element (ME) calculation interfaced with parton showering, and with a fixed-order calculation of a $\PW$ boson and one jet ($\PW$+1-jet) at next-to-NLO (NNLO).

\section{The CMS detector}
\label{cms}

The central feature of the CMS apparatus is a superconducting solenoid of 6\unit{m} internal diameter, providing a magnetic field of 3.8\unit{T}. Within the solenoid volume are a silicon pixel and strip tracker, a lead tungstate crystal electromagnetic calorimeter (ECAL), and a brass and scintillator hadron calorimeter (HCAL), each composed of a barrel and two endcap sections. Forward calorimeters extend the $\eta$ coverage provided by the barrel and endcap detectors. Muons are measured in the range $\abs{\eta} < 2.4$, in gas-ionization detectors with detection planes made using three technologies: drift tubes, cathode strip chambers, and resistive-plate chambers that are embedded in the steel flux-return yoke outside the solenoid. Matching muons to tracks measured in the silicon tracker results in a relative $\pt$ resolution for muons with $20 <\pt < 100\GeV$ of 1.3--2.0\% in the barrel and better than 6\% in the endcaps. The \pt resolution in the barrel is better than 10\% for muons with \pt up to 1\TeV~\cite{Chatrchyan:2012xi}.

The CMS experiment uses a two-level trigger system~\cite{Khachatryan:2016bia}. The first level of the CMS trigger system, composed of custom hardware processors, uses information from the calorimeters and muon detectors to select the most interesting events. The high-level trigger processor farm further decreases the event rate from around 100\unit{kHz} to a rate of around 1\unit{kHz}, before data storage.

A more detailed description of the CMS detector, together with a definition of the coordinate system used and the relevant kinematic variables, can be found in Ref.~\cite{Chatrchyan:2008aa}.

\section{Simulated samples and theoretical calculations}
\label{samples}

Leptonic $\PW$ boson decays are characterized by an energetic isolated lepton, and a neutrino giving rise to significant missing transverse energy $\MET$ in the detector. Background processes with final states similar to the $\PW$+jets signal are Drell--Yan ($\PZ/\gamma^{*}$+jets), $\ttbar$, single top quark, $\PW\PW$/$\PW\PZ$/$\PZ\PZ$+jets (diboson), and QCD multijet production. Signal and background processes are predicted from MC simulations, and their samples are produced and fully reconstructed using a simulation of the CMS detector based on \GEANT{}4 (v9.4p03)~\cite{Agostinelli:2002hh}, except for the QCD multijet background, which is estimated from control data samples. The processes of the $\PW$+jets signal and the $\PZ/\gamma^{*}$+jets background are generated by \MADGRAPH{}5\_a\MCATNLO (v5.2.2.2)~\cite{Alwall:2014hca} with an NLO calculation. The {\sc FxFx} jet merging scheme~\cite{Frederix:2012ps} is used in the \MADGRAPH{}5\_a\MCATNLO generator. The $\ttbar$ background is generated at NLO with \POWHEG (v2.0)~\cite{Nason:2004rx,Frixione:2007vw,Alioli:2010xd}. The single top quark background processes are simulated either with \POWHEG (v1.0)~\cite{Re:2010bp} or with \MADGRAPH{}5\_a\MCATNLO depending on the particular channel. Among the diboson background processes, the $\PW\PW$ production is generated with \POWHEG (v2.0)~\cite{Melia:2011tj}, while the $\PW\PZ$ and $\PZ\PZ$ productions are generated using \PYTHIA{}8 (v8.212)~\cite{Sjostrand:2006za, Sjostrand:2014zea}. The signal and background simulated samples, except for diboson production, are interfaced with \PYTHIA{}8 for parton showering and hadronization. The CUETP8M1 tune~\cite{Khachatryan:2015pea} was used in \PYTHIA{}8. The NNPDF~2.3 LO PDF~\cite{Ball:2010de,Ball:2011mu} and the NNPDF~3.0 NLO PDF~\cite{Ball:2014uwa} are used to generate background processes, where the former is used in \PYTHIA{}8.
The simulated processes include the effect of additional pp collisions in the same or adjacent bunch crossings (pileup). The pileup contribution is simulated as additional minimum bias events superimposed on the primary simulated events based on a distribution of the number of interactions per bunch crossing with an average of about 11 collisions, which is reweighted to match that observed in data.

The measured $\PW$+jets differential cross sections are compared to two multileg ME calculations, an NLO prediction corresponding to the generated $\PW$+jets signal process described above, and an LO prediction. The LO prediction is generated with \MADGRAPH{}5\_a\MCATNLO interfaced with \PYTHIA{}8 for parton showering and hadronization. The ME calculation includes the five processes $\Pp\Pp \rightarrow \PW$+$N\text{-jets}$, $N=0\dots4$ and it is matched to the parton showering using the $\kt$-MLM~\cite{Alwall:2007fs,Alwall:2008qv} scheme with the merging scale set at 19\GeV. The NLO prediction is generated with \MADGRAPH{}5\_a\MCATNLO interfaced with \PYTHIA{}8 for parton showering and hadronization. The {\sc FxFx} merging scheme is used with a merging scale parameter set to~30\GeV. This prediction has an NLO accuracy for $\Pp\Pp \rightarrow \PW$+$N\text{-jets}$, $N=0,\,1,\, 2$, and LO accuracy for $N=3,\, 4$. In the rest of the paper, LO and NLO predictions by \MADGRAPH{}5\_a\MCATNLO are referred to as LO MG\_aMC (MG\_aMC + PY8 ($\leq$ 4j LO + PS) in the figure legends) and NLO MG\_aMC FxFx (MG\_aMC FxFx + PY8 ($\leq$ 2j NLO + PS) in the figure legends), respectively. The NNPDF~3.0 LO (NLO) is used for the ME calculation in LO (NLO FxFx) MG\_aMC prediction, while the NNPDF~2.3 LO PDF is used in the parton showering and hadronization with \PYTHIA{}8 using the CUETP8M1 tune. The total cross section for the LO MG\_aMC prediction is normalized to NNLO cross section calculated with $\FEWZ$ (v3.1)~\cite{Gavin:2012sy}. The measured differential cross sections are also compared to the fixed-order calculation based on the $N$-jettiness subtraction scheme ($\rm N_{jetti}$) at NNLO for $\PW$+1-jet production~\cite{Boughezal:2015dva,Boughezal:2016dtm}. The comparison is made for the measured distributions of the leading jet $\pt$ and $|y|$, $\HT$, $\Delta\phi(\text{$\PGm$, $j_{1}$})$, and $\Delta R(\text{$\PGm$, closest jet})$ for events with one or more jets. The NNPDF~3.0 NNLO PDF is used in this calculation. To account for nonperturbative effects in the NNLO, the predictions with and without multiple parton interactions and hadronization are computed with LO MG\_aMC interfaced with \PYTHIA{}8. The value of this multiplicative correction applied to the NNLO calculation is mostly within the range of 0.93--1.10. The effect of final-state radiation (FSR) from the muon on the NNLO prediction is estimated to be less than 1\%.

\section{Particle reconstruction and identification}
\label{objectreco}

The final state candidates in $\PW(\PGm\PGn)$ decays are identified and reconstructed with the particle-flow~(PF) algorithm~\cite{Sirunyan:2017ulk}, which combines information from all the CMS subdetectors.

Muon PF candidates are identified as tracks in the muon system that are matched to tracks reconstructed in the inner tracking system. Muon candidates are required to have $\pt >25\GeV$ inside the acceptance of $\abs{\eta}< 2.4$ and to satisfy the tight identification criteria~\cite{Chatrchyan:2012xi}. In addition, an isolation requirement, known as the combined relative PF isolation requirement, is applied to suppress the contamination from muons contained in jets:
\begin{equation}
\ifthenelse{\boolean{cms@external}}
{
\begin{split}
I = & \\ \left[\sum^{\text{Ch.had.}}\pt \right. & \left. + \,
 \text{max}(0, \sum^{\text{N.had.}}\pt + \sum^{\text{EM}}\pt - 0.5\sum^{\text{PU}}\pt)\right]/\pt^{\PGm} \\
& \leq 0.15,
\end{split}
}
{
I = \left[\sum^{\text{Ch.had.}}\pt + \text{max}(0, \sum^{\text{N.had.}}\pt + \sum^{\text{EM}}\pt - 0.5\sum^{\text{PU}}\pt)\right]/\pt^{\PGm} \leq 0.15,
}
\end{equation}
where the sums run over the charged hadrons (Ch.had.), neutral hadrons (N.had.), photons (EM), and charged particles from pileup (PU), inside a cone of radius $\Delta R \leq 0.4$ around the muon direction. This isolation requirement includes a correction for pileup effects, which is based on the scalar $\pt$ sum of charged particles not associated with the primary vertex ($0.5\sum^{\text{PU}}\pt$) in the isolation cone, where the factor 0.5 corresponds to an approximate average ratio of neutral to charged particles and has been measured in jets in Ref.~\cite{Sirunyan:2017ulk}. The small differences in muon identification, isolation, and trigger efficiencies between data and simulated processes are compensated by applying corrections to the simulated events.

Hadronic jets are reconstructed from the PF candidates by using the anti-$\kt$ clustering algorithm~\cite{Cacciari:2008gp}, as implemented in the \FASTJET package~\cite{Cacciari:2011ma}, with a distance parameter of $R = 0.4$. Reconstructed jet energies are corrected by using $\pt$- and $\eta$-dependent corrections to account for the following effects: nonuniformity and nonlinearity of the ECAL and HCAL energy response to neutral hadrons, the presence of extra particles from pileup interactions, the thresholds used in jet constituent selection, reconstruction inefficiencies, and possible biases introduced by the clustering algorithm. Jet energy corrections are derived based on the measurement of the $\pt$ balance in dijet and $\gamma$+$\text{jet}$ events~\cite{Khachatryan:2016kdb}. A residual $\pt$- and $\eta$-dependent calibration is applied to the jets in data to correct for the small differences between data and simulation. The jets in simulated events are smeared by an $\eta$-dependent factor to account for the difference in energy resolution between data and simulation~\cite{Khachatryan:2016kdb}. Jets are required to have $\pt >30\GeV$ inside the acceptance of $|y| < 2.4$ and a spatial separation of $\Delta R > 0.4$ from muon candidates. Additional loose selection criteria are applied to each event to suppress nonphysical jets~\cite{CMS:2017wyc}. A number of vertexing-related and jet-shower-shape-related input variables are combined into a boosted decision tree yielding a single discriminator for the identification of pileup jets~\cite{CMS:2017wyc}. The contribution from pileup jets is reduced by applying a selection on the discriminator that has been optimized to minimize the dependency on the number of reconstructed vertices.

In leptonic $\PW$ boson decays, neutrinos pass through the detector without interacting and their presence is a fundamental tool for the discrimination of $\PW$ boson events from the main backgrounds. The presence of a neutrino is indicated by $\MET$ information in the event. The missing transverse momentum vector $\ptvecmiss$ is the projection on the plane perpendicular to the beams of the negative vector sum of the momenta of all PF candidates reconstructed in the event. $\MET$ is defined as the magnitude of the $\ptvecmiss$ vector and it is a measure of the $\pt$ of particles leaving the detector undetected. For the reconstructed $\MET$, the vector $\pt$ sum of particles that were clustered as jets is replaced by the vector $\pt$ sum of the jets including the jet energy corrections. Moreover, a set of individual $\MET$ filters are applied to veto events with anomalous $\MET$ due to various subdetector malfunctions and algorithmic errors~\cite{CMS-PAS-JME-16-004}.

\section{Event selection}
\label{Weventselection}

The $\PW(\PGm\PGn)$+jets events are required to have exactly one muon and one or more jets. Data events are retained if they pass an online trigger requirement with a muon reconstructed in the online system with $\pt > 20\GeV$, while the simulated events are required to pass an emulation of the trigger requirement. Muons are required to satisfy the muon selection criteria including $\pt> 25\GeV $ and acceptance of $|\eta| < 2.4$, and jets are required to satisfy the jet selection criteria with $\pt > 30\GeV$ and acceptance of $|y| < 2.4$, as described in Section~\ref{objectreco}. Events with additional muon PF candidates, which are not necessarily subject to the muon identification and isolation criteria, with $\pt> 15\GeV $ and $|\eta| < 2.4$ are removed. Events are further required to be in the transverse mass peak region for $\PW$ bosons, defined by $m_{\text{T}} > 50\GeV$, where $m_{\text{T}}$ is the invariant transverse mass between the muon and the $\ptvecmiss$ variable, which can be formulated as $m_{\text{T}}(\PGm,\ptvecmiss) \equiv \sqrt{\smash[b]{2\pt^\PGm \MET(1-\cos\Delta\phi)}}$, where $\Delta\phi$ is the difference in azimuthal angle between the direction of the muon momentum $\ptvec^{\PGm}$ and $\ptvecmiss$. The $m_{\text{T}}$ variable selection operates as a discriminator against non-$\PW$ final states, such as QCD multijet background, that have a lepton candidate and nonzero $\ptvecmiss$, but a relatively low value of $m_{\text{T}}$. For the analysis of the $\Delta R(\text{$\PGm$, closest jet})$ distribution, jets in the event are required to have $\pt > 100\GeV $, with the leading jet $\pt > 300\GeV$. This selection results in a boosted topology, where two jets recoil against each other and one of them can lose a significant amount of energy to the decay products of the emitted real $\PW$ boson.

The primary background to the $\PW$+jets production at high jet multiplicities (i.e., four or more) is $\ttbar$ production. The contamination from $\ttbar$ events is reduced by applying a \PQb-quark tag veto to the events that contain one or more \PQb-tagged jets. For this veto, the combined secondary vertex tagger~\cite{Chatrchyan:2012jua} is used as the \PQb~tagging algorithm with medium discrimination working point~\cite{CMS-PAS-BTV-15-001} corresponding to the misidentification probability of approximately 1\% for light-flavor jets with $\pt> 30\GeV$. After the implementation of the \PQb~tag veto, the expected contributions from the background processes and the observed data are given in Table~\ref{events} as a function of the jet multiplicity. For jet multiplicities of 1--6, the \PQb~tag veto rejects 71--88\% of the predicted $\ttbar$ background and 5--29\% of the $\PW$+jets signal. Differences in the data and simulation \PQb~tagging efficiencies and mistagging rates are corrected by applying data-to-simulation scale factors~\cite{CMS-PAS-BTV-15-001}.

\begin{table*}
\centering
\topcaption{Numbers of events in simulation and data as a function of the exclusive jet multiplicity after the implementation of \PQb~tag veto. The processes included are: $\PW\PW$, $\PW\PZ$, and $\PZ\PZ$ diboson ($\PV\PV$), QCD multijet, single top quark (Single t), $\PZ/\gamma^{*}$+jets Drell--Yan (DY+jets), $\ttbar$, and $\PW(\PGm\PGn)$+jets signal processes. The QCD multijet background is estimated using control data samples. The $\ttbar$ background is scaled as discussed in Section~\ref{WdataMC}.}
\begin{scotch}{l|rrrrrrr}
$N_{\text{jets}} $ &  \multicolumn{1}{c}{$ 0 $} & \multicolumn{1}{c}{$ 1 $} & \multicolumn{1}{c}{$ 2 $} & \multicolumn{1}{c}{$ 3 $} & \multicolumn{1}{c}{$ 4 $} & \multicolumn{1}{c}{$ 5 $} & \multicolumn{1}{c}{$ 6 $} \\ \hline
$\PV\PV$        & 4\,302 \x & 1\,986 \x & 774 \x & 205 \x & 45 \x & 10 \x & 2 \x \\
QCD multijet        & 205\,800 \x & 75\,138 \x & 12\,074 \x & 2\,556 \x & 612 \x & 53 \x & 5 \x \\
Single $\PQt$        & 3\,392 \x & 5\,484 \x & 3\,277 \x & 1\,194 \x & 317 \x & 83 \x & 19 \x \\
DY+jets        & 520\,653 \x & 69\,660 \x & 14\,666 \x & 3\,041 \x & 643 \x & 133 \x & 33 \x \\
$\ttbar$         & 1\,663 \x & 4\,901 \x & 8\,084 \x & 6\,170 \x & 3\,152 \x & 1\,152 \x & 319 \x \\
$\PW(\PGm\PGn)$+jets       & 12\,171\,400 \x & 1\,601\,858 \x & 326\,030 \x & 64\,484 \x & 11\,736 \x & 2\,072 \x & 404 \x \\
\hline
Total & 12\,907\,210 \x & 1\,759\,027 \x & 364\,905 \x & 77\,650 \x & 16\,505 \x & 3\,503 \x & 782 \x \\
\hline
Data          & 12\,926\,230 \x & 1\,680\,182 \x & 349\,480 \x & 73\,817 \x & 16\,866 \x & 3\,964 \x & 909 \x \\
\end{scotch}
\label{events}
\end{table*}

\section{Data-to-simulation comparisons}
\label{WdataMC}

The goal of this analysis is to measure the differential cross sections characterizing the production of a $\PW$ boson and associated jets as functions of several kinematic and angular observables with 13\TeV data. We first compare data with simulated processes at the reconstruction level for some of the observables that are used for the cross section measurement. Signal and background processes in these comparisons are simulated with the event generators described in Section~\ref{samples}, with the exception of the QCD multijet background, which is estimated using a data control region with an inverted muon isolation requirement. In the data control region, the muon misidentification rate for multijet processes is estimated in a sideband region with $m_{\text{T}} < 50\GeV$ and the multijet distribution shape template is extracted in a region with $m_{\text{T}} > 50\GeV$. The muon misidentification rate is then used to rescale the multijet shape template. This estimation method was used in the measurement of the $\PW$+jets production cross section at 7\TeV, and it is described in detail in Ref.~\cite{Khachatryan:2014uva}.

At high jet multiplicities, where the $\PW$+jets signal is less dominant, the accuracy of the background modeling becomes more important, especially for the $\ttbar$ production process. We created a $\ttbar$-enriched control sample by requiring two or more \PQb-tagged jets. The purity of this $\ttbar$ control sample increases towards higher jet multiplicities and ranges between 79--96\% for jet multiplicities of 2--6. The differences between data and simulation observed for jet multiplicities of 2--6 in the $\ttbar$ control region are expressed in terms of $\ttbar$ data-to-simulation scaling factors that range between 0.75 and 1.15. The $\ttbar$ background events are scaled by these factors in all the reconstructed-level and unfolded distributions presented in this paper for events with jet multiplicities of 2--6.

\begin{figure}[!htbp]
\begin{center}
\includegraphics[width=0.40\textwidth]{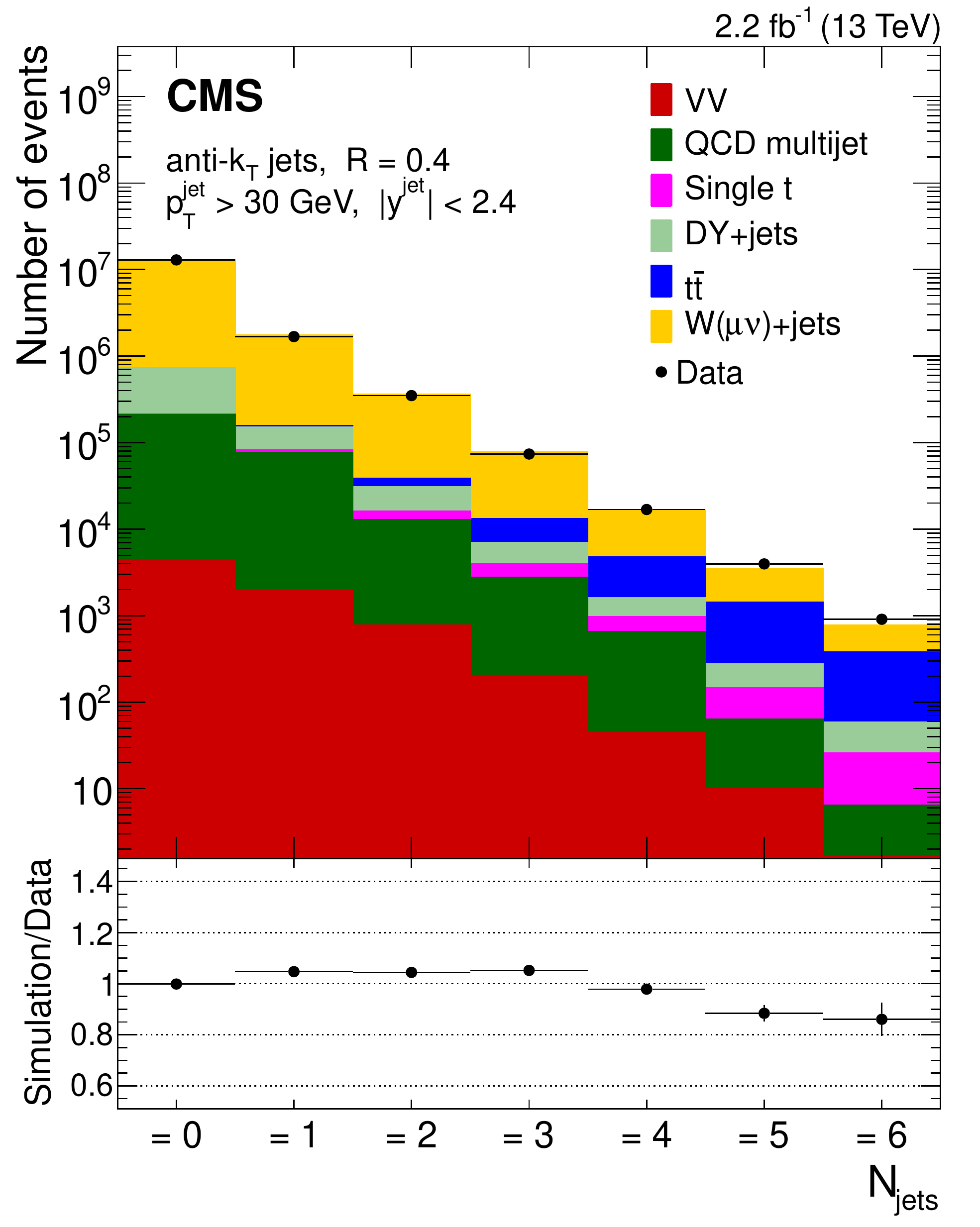}
\end{center}
\caption{Data-to-simulation comparison as a function of the jet multiplicity. The processes included are listed in Table~\ref{events}. The QCD multijet background is estimated using control samples in data. The $\ttbar$ background is scaled as discussed in Section~\ref{WdataMC}. The error bars in the ratio panel represent the combined statistical uncertainty of the data and simulation.}
\label{mult}
\end{figure}

\begin{figure*}[!htbp]
\begin{center}
\includegraphics[width=0.40\textwidth]{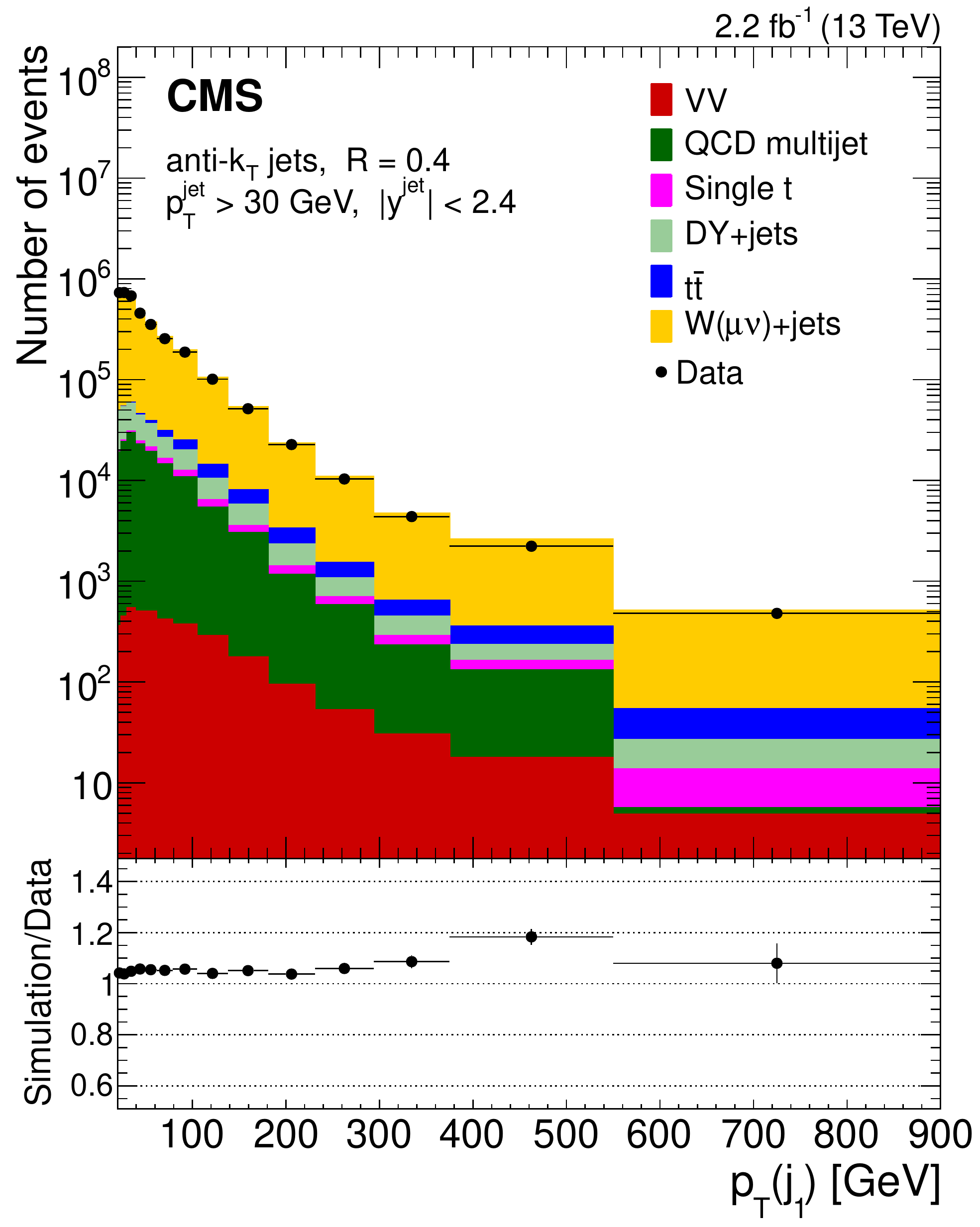}
\hspace{0.05\textwidth}
\includegraphics[width=0.40\textwidth]{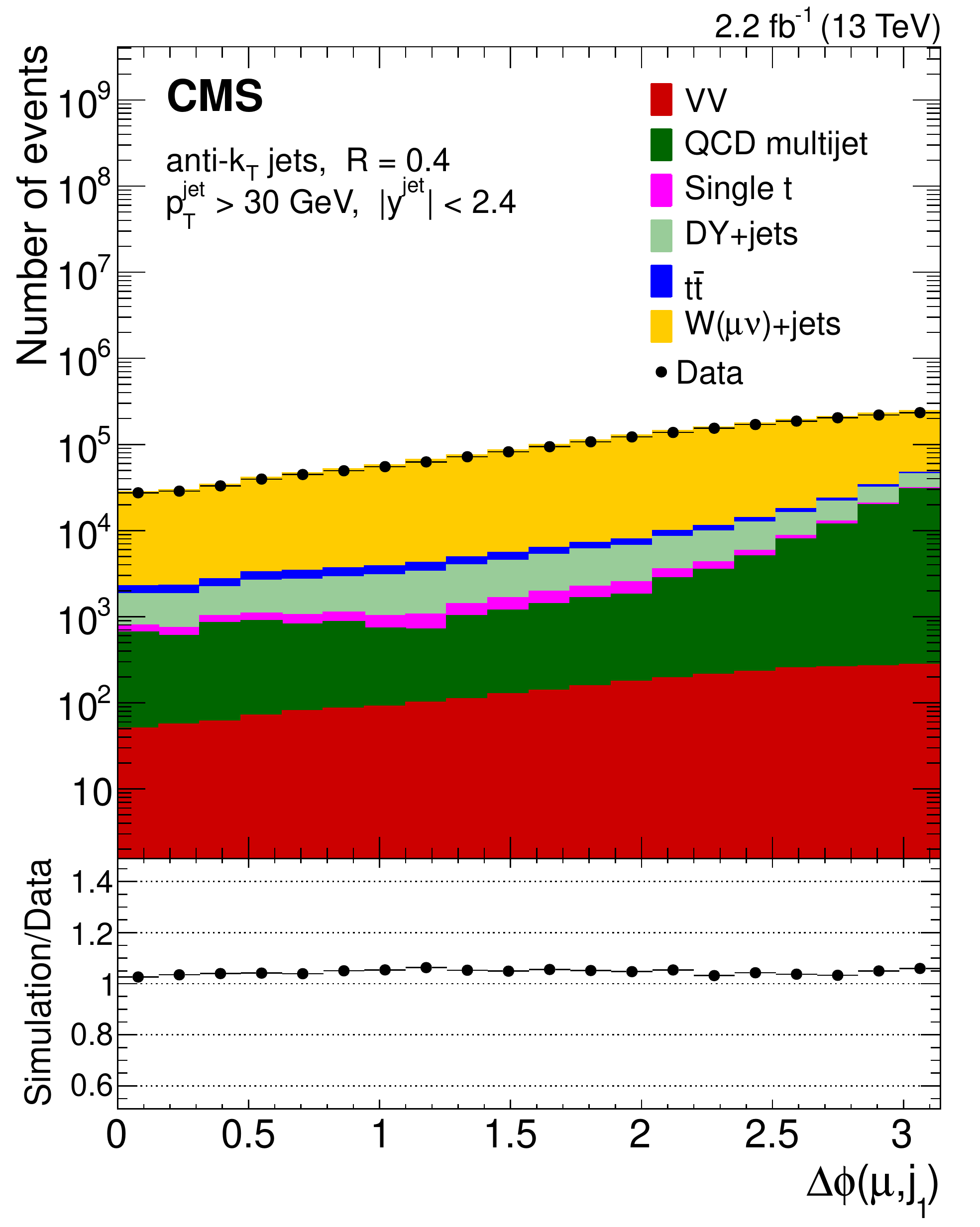}\\
\end{center}
\caption{Data-to-simulation comparison as functions of the leading jet $\pt$ (left) and $\Delta\phi(\text{$\PGm$, $j_{1}$})$ between the muon and the leading jet (right) for one jet inclusive production. The processes included are listed in Table~\ref{events}. The QCD multijet background is estimated using control samples in data. The $\ttbar$ background is scaled as discussed in Section~\ref{WdataMC}. The error bars in the ratio panel represent the combined statistical uncertainty of the data and simulation.}
\label{pt1}
\end{figure*}

The comparison of reconstructed distributions for data and simulated processes is shown in Fig.~\ref{mult} for the jet multiplicity. The $\pt$ distribution of the leading jet and the azimuthal correlation between the muon and the leading jet in events with at least one jet are shown in Fig.~\ref{pt1}. For each reconstructed distribution, the ratio of the sum of the simulated processes from signal and backgrounds to the data is presented to quantify possible disagreements (the corresponding error bars represent the statistical uncertainties stemming from both data and simulation). The data-to-simulation agreement is on the 5\% level in almost all regions.

\section{Unfolding procedure}
\label{unfolding}

The measured distributions are obtained by subtracting the estimated backgrounds from data, where the $\ttbar$ events are scaled as discussed in Section~\ref{WdataMC}. These background-subtracted distributions are corrected to the stable-particle level using an unfolding procedure for the detector effects, such as efficiency and resolution, migration of events between neighboring bins, and measurement acceptance, in order to obtain a measured cross section directly comparable to the theoretical predictions. This unfolding procedure is applied separately to each measured differential cross section and includes corrections for the trigger and the muon selection efficiencies. This procedure is performed using the method of D'Agostini iteration with early stopping~\cite{DAgostini:1994zf,Richardson:72,Lucy:1974yx} that is implemented in the statistical analysis toolkit {\sc RooUnfold}~\cite{Adye:2011gm}. The unfolding procedure is described briefly below.

A response matrix, which defines the event migration probability between the particle-level and reconstructed quantities, is constructed using generator and reconstruction levels of the NLO MG\_aMC FxFx $\PW$+jets simulated sample, respectively.
At the generator level, the events are required to pass the same kinematic selection used at the reconstruction level as described in Section~\ref{Weventselection}, including the requirements on muon $\pt$ and $|\eta|$, jet $\pt$ and $|y|$, and $m_{\text{T}}$. The generated values refer to the stable leptons, a muon and a neutrino, from the decay of the $\PW$ boson and to jets built from stable particles excluding neutrinos, using the same algorithm as for the measurement. Particles are considered stable if their decay length $c\tau$ is greater than 1\unit{cm}. The muons are 'dressed' by recombining the bare muons and all of the radiated photons in a cone of $\Delta R < 0.1$ around the muon to account for the FSR effects, which may have a significant contribution. At the generator level, the $m_{\text{T}}$ of the $\PW$ boson is calculated using the dressed muon and the neutrino. The response matrix, the reconstructed-level distribution and the corresponding generator-level distribution contain information about the migration probability between the reconstructed and generated quantities as well as the information used to determine the effect of inefficiencies. Events that pass the reconstructed-level selection but are absent from the generator-level selection due to migrations across neighboring bins are estimated, and these events are subtracted from the measured distribution. The unfolding procedure is regularized by choosing the number of iterations in the D'Agostini method. The iteration is optimized by folding the unfolded distributions corrected with the response matrix and comparing to the initial measured distributions. The number of iterations is determined when the distribution obtained by folding the unfolded distribution becomes compatible with the initial measured distribution based on a $\chi^2$ comparison. A minimum of four iterations is required to avoid biasing the unfolded results towards the simulated sample used to construct the response matrix.

\section{Systematic uncertainties}
\label{systematics}

The dominant source of systematic uncertainty in the differential cross sections, which are measured as functions of the variables involving jet kinematics, is generally the jet energy scale (JES) uncertainty. It is estimated by varying the corresponding jet energy corrections by one standard deviation as described in Ref.~\cite{Khachatryan:2016kdb}. This JES uncertainty is $\pt$- and $\eta$-dependent, and it can be as large as 7\% in the leading jet $\pt$ and $|y|$ measurements. Moreover, this uncertainty increases with the number of reconstructed jets and ranges between 1--25\% for jet multiplicities of 1--6. These JES uncertainties are propagated to the calculation of $\MET$.

As described in Section~\ref{samples}, scale factors are applied to correct for the data-to-simulation difference in jet energy resolution. The uncertainty in these factors is assessed by scaling the energies of the jets in the $\PW$+jets sample with two additional sets of scale factors that correspond to varying the factors up and down by one sigma and evaluating the impact of these new sets. The resulting uncertainty is of the order of 1\%.

The cross sections of the background processes are varied within their theoretical uncertainties. The cross section of the largest background contribution in the higher jet multiplicities, coming from $\ttbar$ production process, is varied by 6\%, which is a sum in quadrature of the theoretical uncertainty in the $\ttbar$ cross section at NNLO accuracy and the statistical uncertainties in the factors used to scale the simulated events to the observed data in the $\ttbar$-enriched control region described in Section~\ref{WdataMC}. The diboson backgrounds are simultaneously varied up and down by their theoretical cross section uncertainties of scale and PDF from \MCFM (v6.6)~\cite{Campbell:2010ff} as 6\% for $\PW\PW$ and 7\% for $\PW\PZ$ and $\PZ\PZ$. The $\PZ/\gamma^{*}$+jets background is varied by scale and PDF uncertainties of 4\% obtained from \FEWZ (v3.1). The single top quark processes are varied by scale, PDF, and $\alpha_{s}$ uncertainties of 4\% for the $s$- and $t$-channels and by scale and PDF uncertainties of 6\% for the $\PQt\PW$-channel. The QCD multijet background is estimated using data control regions and has an uncertainty based on the number of events in an inverted muon isolation sample, and in the low-$m_{\text{T}}$ control regions, with nominal and inverted isolation requirements, which are used to calculate the normalization. The uncertainty associated with the estimated multijet background is 0.6--46\% depending on the jet multiplicity.

The uncertainties in the data-to-simulation correction factors of the \PQb~tagging efficiencies are also considered. This systematic uncertainty is assessed by adjusting the scale factors up and down according to their uncertainties. The entire analysis is performed with these variations and the final unfolded results are compared to the results of the standard analysis. The difference is taken to be the systematic uncertainty. The effect of this uncertainty on the measured cross section changes between 0.3 and 12\%, depending on the jet multiplicity.

The uncertainty in the data-to-simulation scale factors for the efficiency of muon selection is the sum in quadrature of the systematic uncertainties in the trigger, identification, and isolation efficiencies and is equal to 1.2\%.

A systematic uncertainty associated with the generator used to build the unfolding response matrix is estimated by weighting the simulated events to agree with the data in $\pt$, $|y|$, $\HT$, $\Delta\phi(\text{$\PGm$, $j_{i}$})$, and $\Delta R(\text{$\PGm$, closest jet})$ distributions and building a reweighted response matrix to unfold the data. The reweighting is done using a finer binning. The difference between the nominal results, and the results unfolded using the reweighted response matrix is taken as the systematic uncertainty associated with the unfolding response matrix. An uncertainty due to the finite number of simulated events used to construct the response matrix is estimated by propagating the uncertainty in the MEs through the unfolding. This uncertainty ranges from 0.1\% to 12\% for jet multiplicities of 1--6 and becomes the dominant source of systematic uncertainty in regions of phase space with low statistical precision especially in the $\Delta R(\text{$\PGm$, closest jet})$ measurement.

The uncertainty in the pileup condition and modeling in simulated samples is assessed by varying the inelastic pp cross section, which is used to estimate the pileup contribution in data, from its central value within its determined uncertainty of ${\pm}5\%$. The resulting uncertainty is of the order of 1\%.

The uncertainty in the integrated luminosity is estimated to be 2.3\% using a method described in Ref.~\cite{CMS-PAS-LUM-15-001}.

\section{Results}
\label{Wresults}

The size of the data sample used in this analysis allows the measurements of cross sections of the $\PW(\PGm\PGn)$+jets process for jet multiplicities up to six and fiducial cross sections as functions of several kinematic observables for up to four inclusive jets.

The measured $\PW$+jets differential cross section distributions are shown here in comparison with the predictions of the multileg NLO MG\_aMC FxFx and multileg LO MG\_aMC tree level $\kt$-MLM event generators, as described in Section~\ref{samples}. Furthermore, the measured cross sections are compared to the fixed-order $\rm N_{jetti}$ NNLO calculation for $\PW$+1-jet production on the leading jet $\pt$ and $|y|$, $\HT$, $\Delta\phi(\text{$\PGm$, $j_{1}$})$, and $\Delta R(\text{$\PGm$, closest jet})$ distributions. The ratios of the predictions to the measurements are provided to make easier comparisons.

Total experimental uncertainties are quoted for the data in the differential cross section distributions. The multileg LO MG\_aMC prediction is given only with its statistical uncertainty. The NLO MG\_aMC FxFx prediction is given with both the statistical and systematic uncertainties. Systematic uncertainties in the NLO MG\_aMC FxFx prediction are obtained by varying the NNPDF~3.0 NLO PDFs and the value of $\alpha_{s}$, and by varying independently the renormalization and factorization scales by a factor of 0.5 and 2. All possible combinations are used in variations of scales excluding only the cases where one scale is varied by a factor of 0.5 and the other one by a factor of 2. The total systematic uncertainty is the squared sum of these uncertainties. The systematic uncertainty due to variation of scale factors for the exclusive jet multiplicity distribution is computed using the method described in Refs.~\cite{Gangal:2013nxa,Stewart:2011cf}. For the NNLO prediction, the theoretical uncertainty includes both statistical and systematic components, where the systematic uncertainty is calculated by varying independently the central renormalization and factorization scales by a factor of 2 up and down, disallowing the combinations where one scale is varied by a factor of 0.5 and the other one by a factor of 2.0.

\begin{figure*}[hp]
\begin{center}
\includegraphics[width=0.45\textwidth]{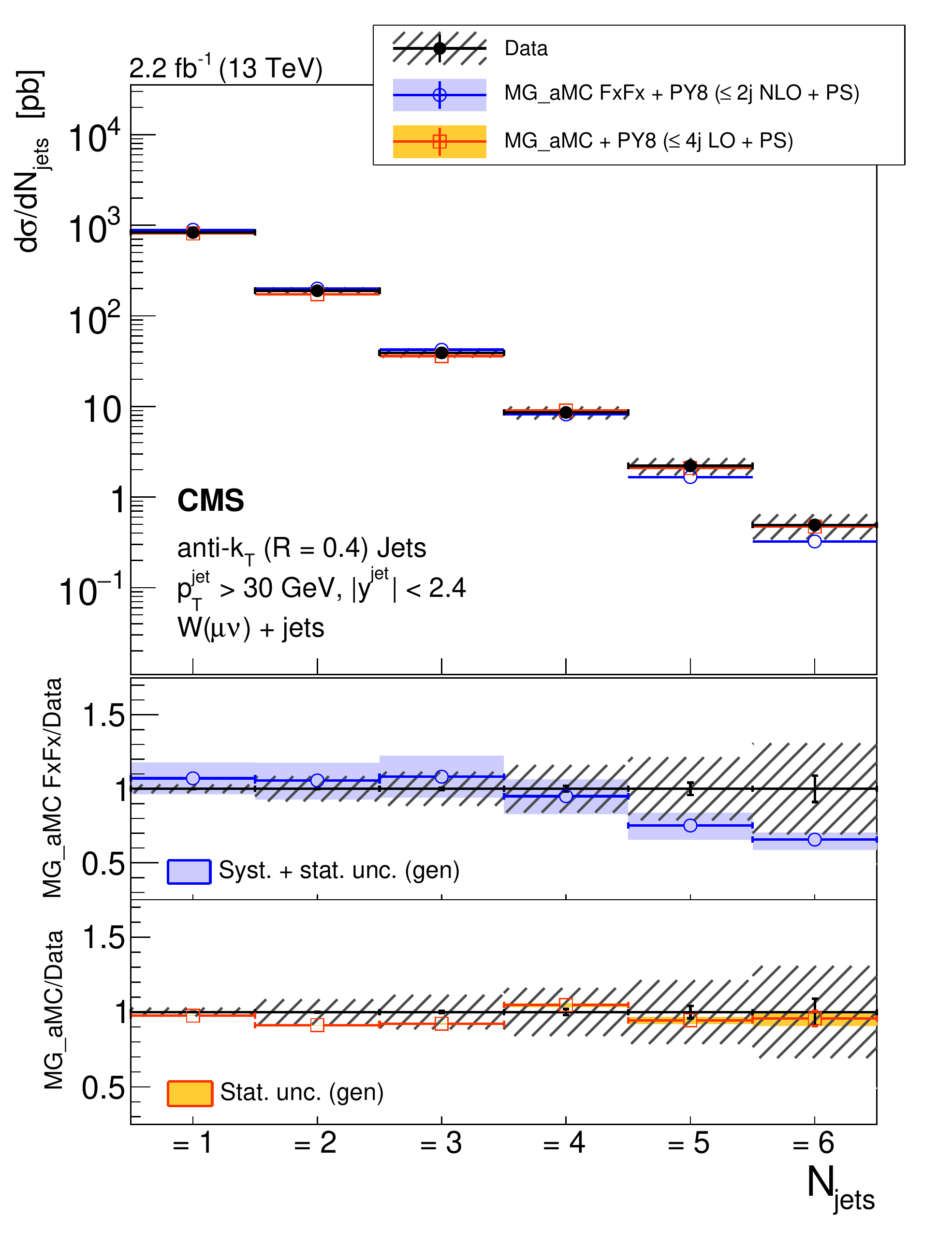}
\includegraphics[width=0.45\textwidth]{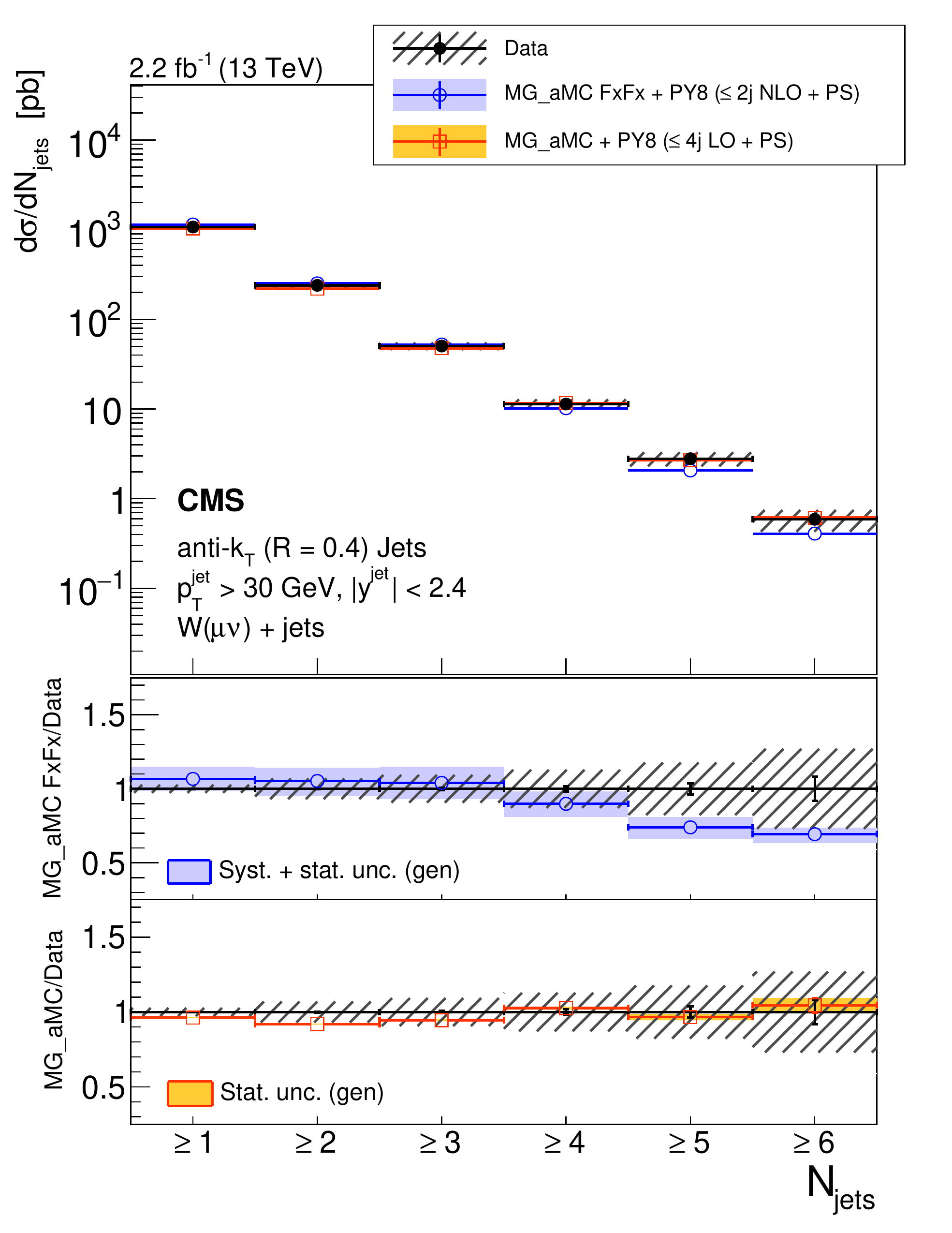}
\caption{Differential cross section measurement for the exclusive (left) and inclusive jet multiplicities (right), compared to the predictions of NLO MG\_aMC FxFx and LO MG\_aMC. The black circular markers with the gray hatched band represent the unfolded data measurement and the total experimental uncertainty. The LO MG\_aMC prediction is given only with its statistical uncertainty. The band around the NLO MG\_aMC FxFx prediction represents its theoretical uncertainty including both statistical and systematic components. The lower panels show the ratios of the prediction to the unfolded data.}
\label{xsec_Njets_nlo}
\end{center}
\end{figure*}

The measured differential cross sections as functions of the exclusive and inclusive jet multiplicities up to 6 jets are compared with the predictions of LO MG\_aMC and NLO MG\_aMC FxFx in Fig.~\ref{xsec_Njets_nlo}. The measured cross sections and the predictions are in good agreement within uncertainties.

\begin{figure*}[!htbp]
\begin{center}
\includegraphics[width=0.45\textwidth]{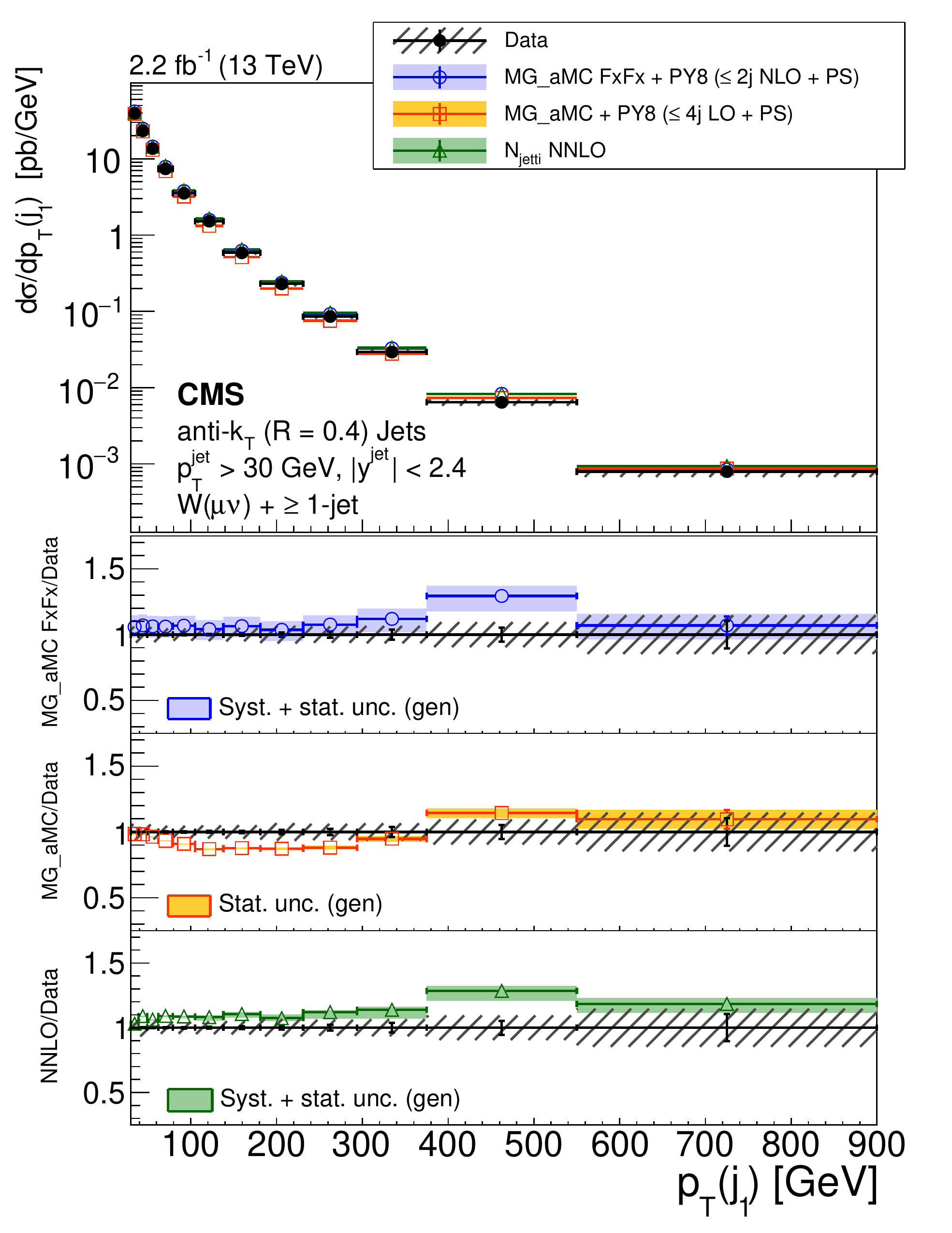}
\includegraphics[width=0.45\textwidth]{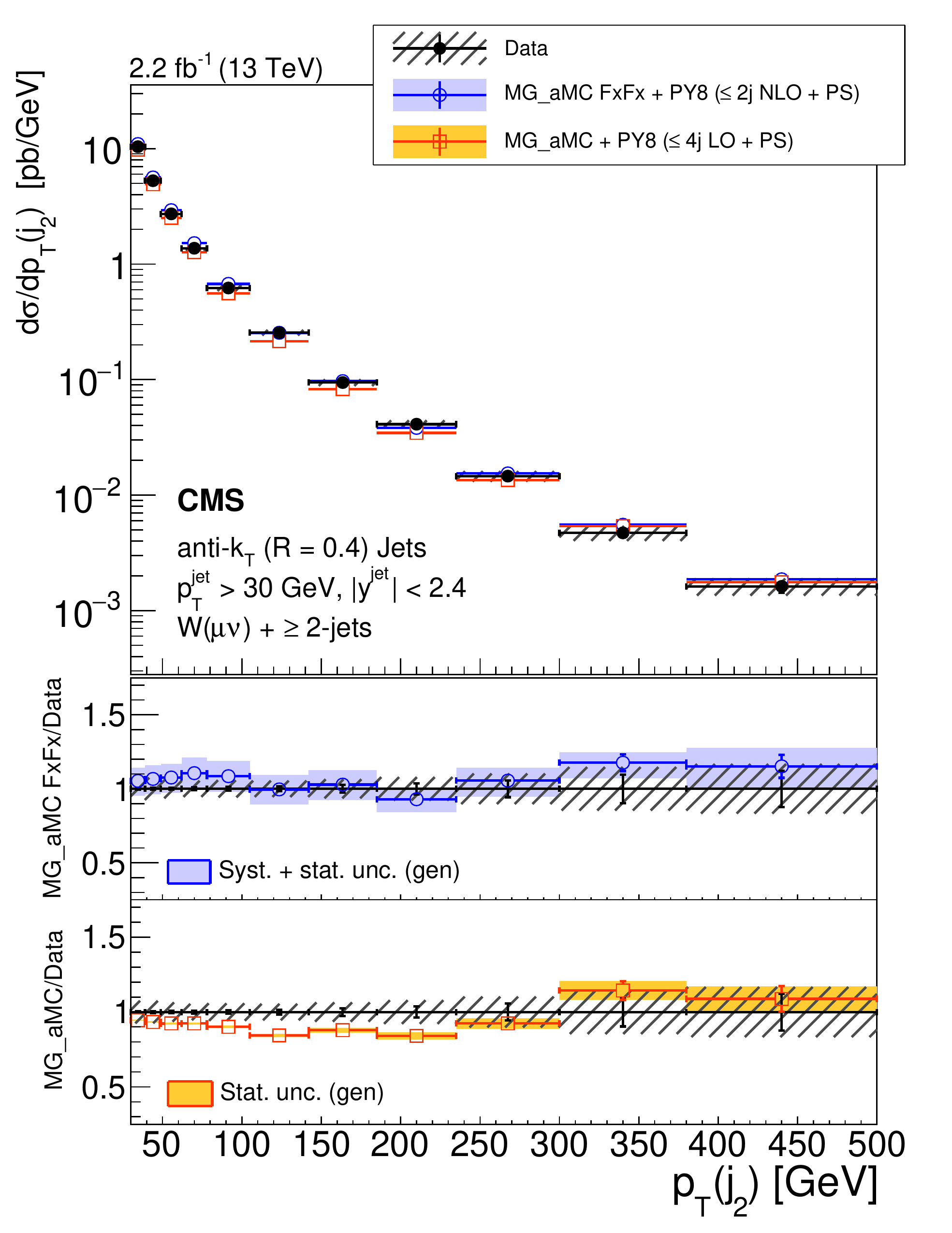}\\
\includegraphics[width=0.45\textwidth]{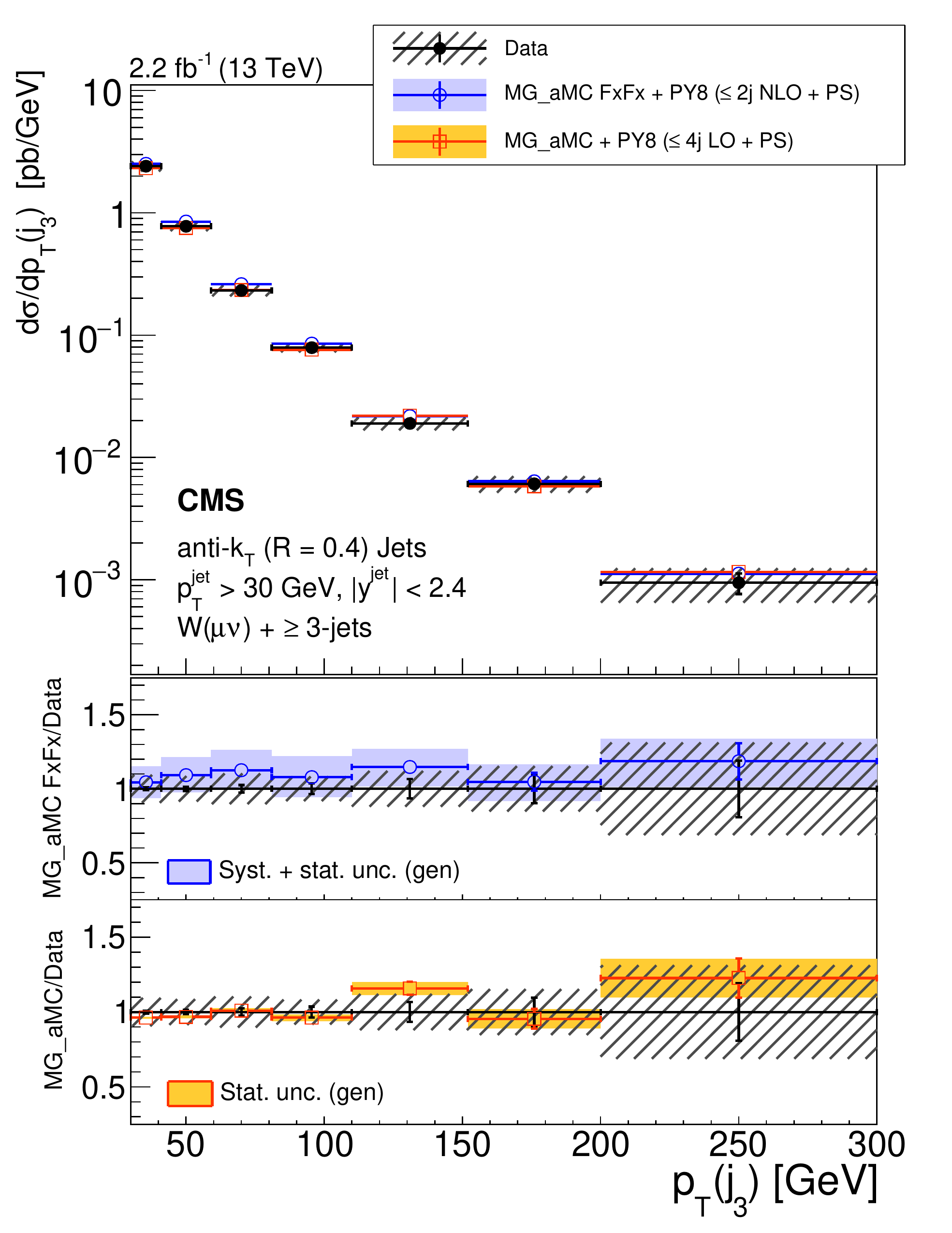}
\includegraphics[width=0.45\textwidth]{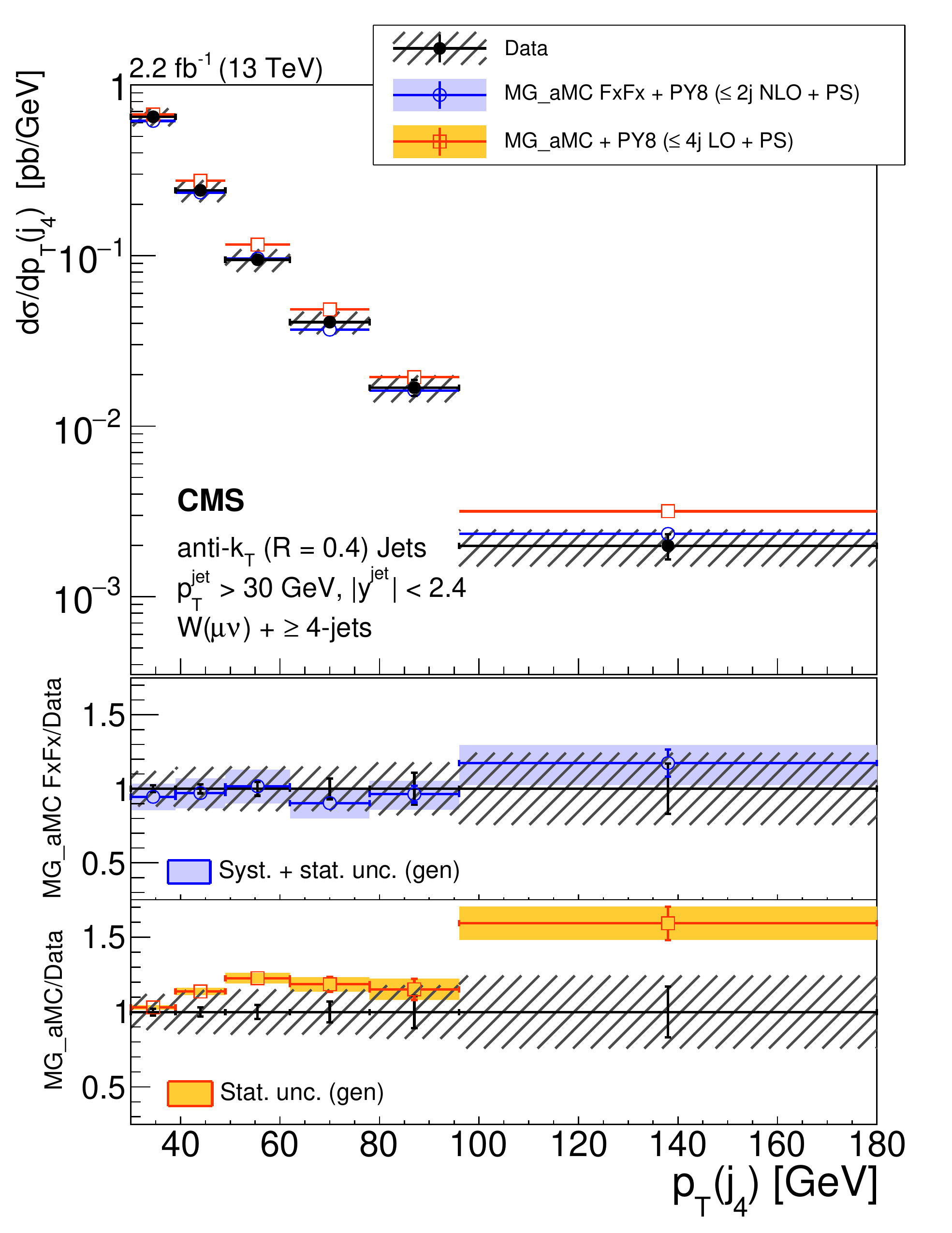}
\caption{Differential cross section measurement for the transverse momenta of the four leading jets, shown from left to right for at least 1 and 2 jets (upper) and for at least 3 and 4 jets (lower) on the figures, compared to the predictions of NLO MG\_aMC FxFx and LO MG\_aMC. The NNLO prediction for $\PW$+1-jet is included in the first leading jet $\pt$. The black circular markers with the gray hatched band represent the unfolded data measurement and the total experimental uncertainty. The LO MG\_aMC prediction is given only with its statistical uncertainty. The bands around the NLO MG\_aMC FxFx and NNLO predictions represent their theoretical uncertainties including both statistical and systematic components. The lower panels show the ratios of the prediction to the unfolded data.}
\label{xsec_JetPt1to3_nlo}
\end{center}
\end{figure*}

\begin{figure*}[!htbp]
\begin{center}
\includegraphics[width=0.45\textwidth]{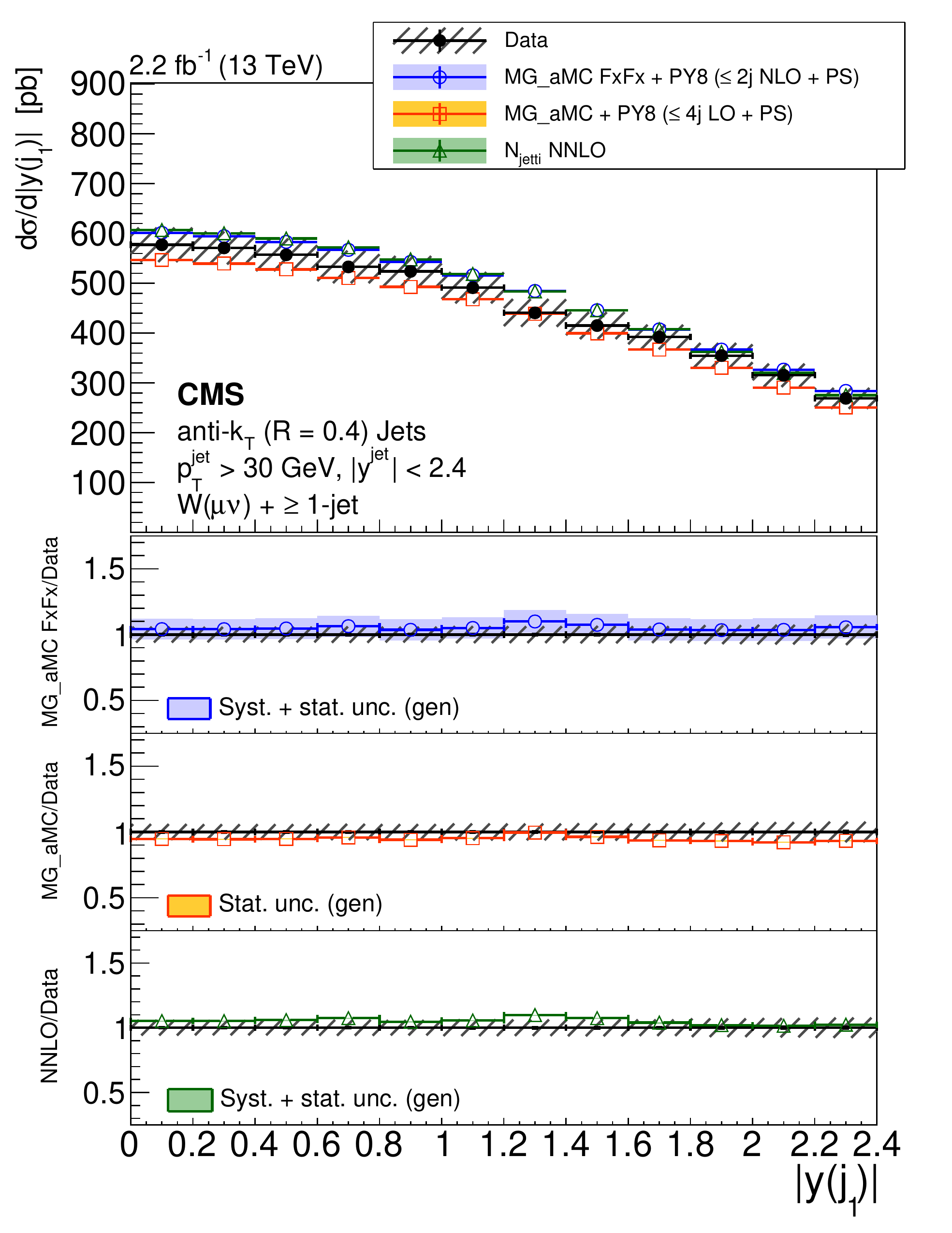}
\includegraphics[width=0.45\textwidth]{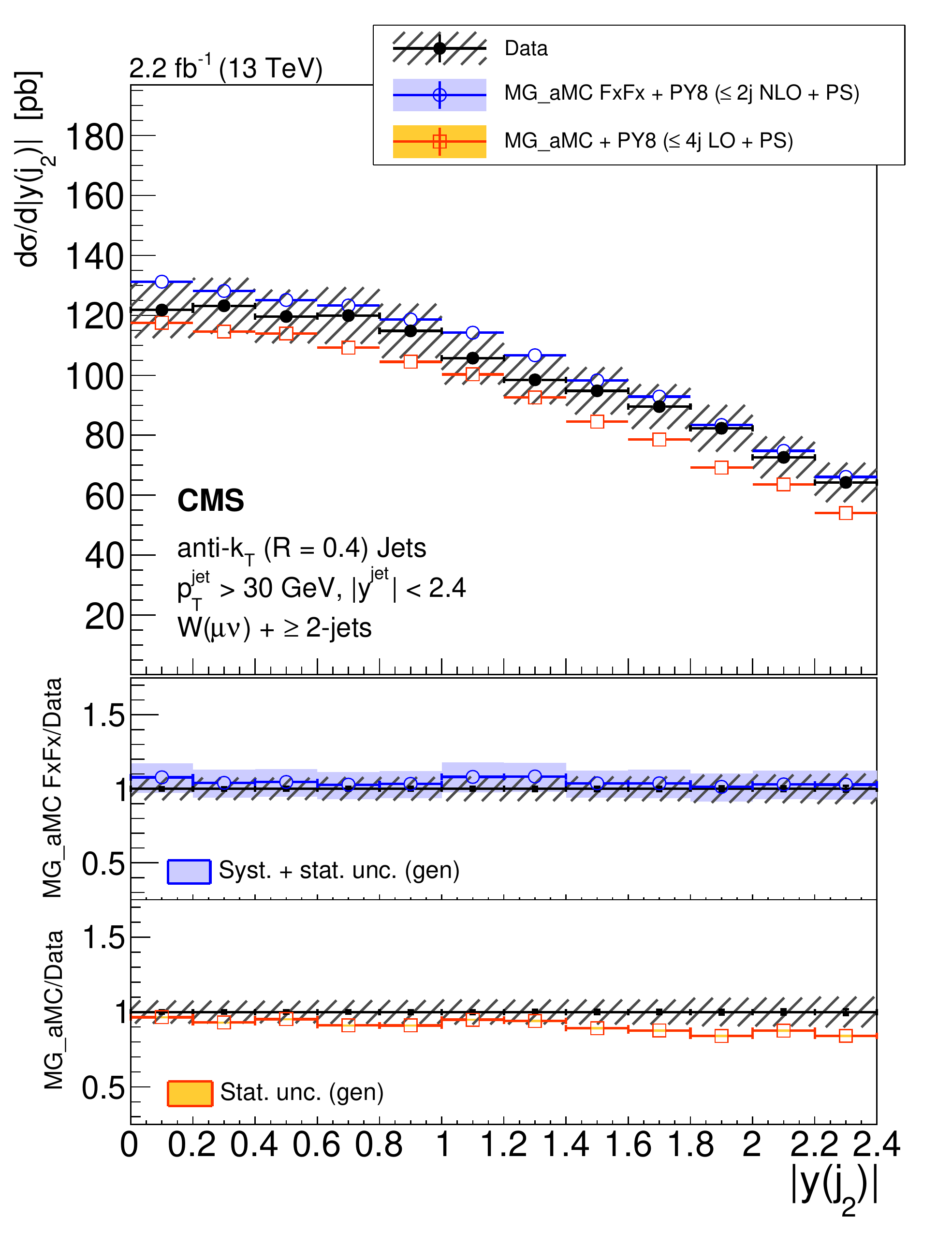}\\
\includegraphics[width=0.45\textwidth]{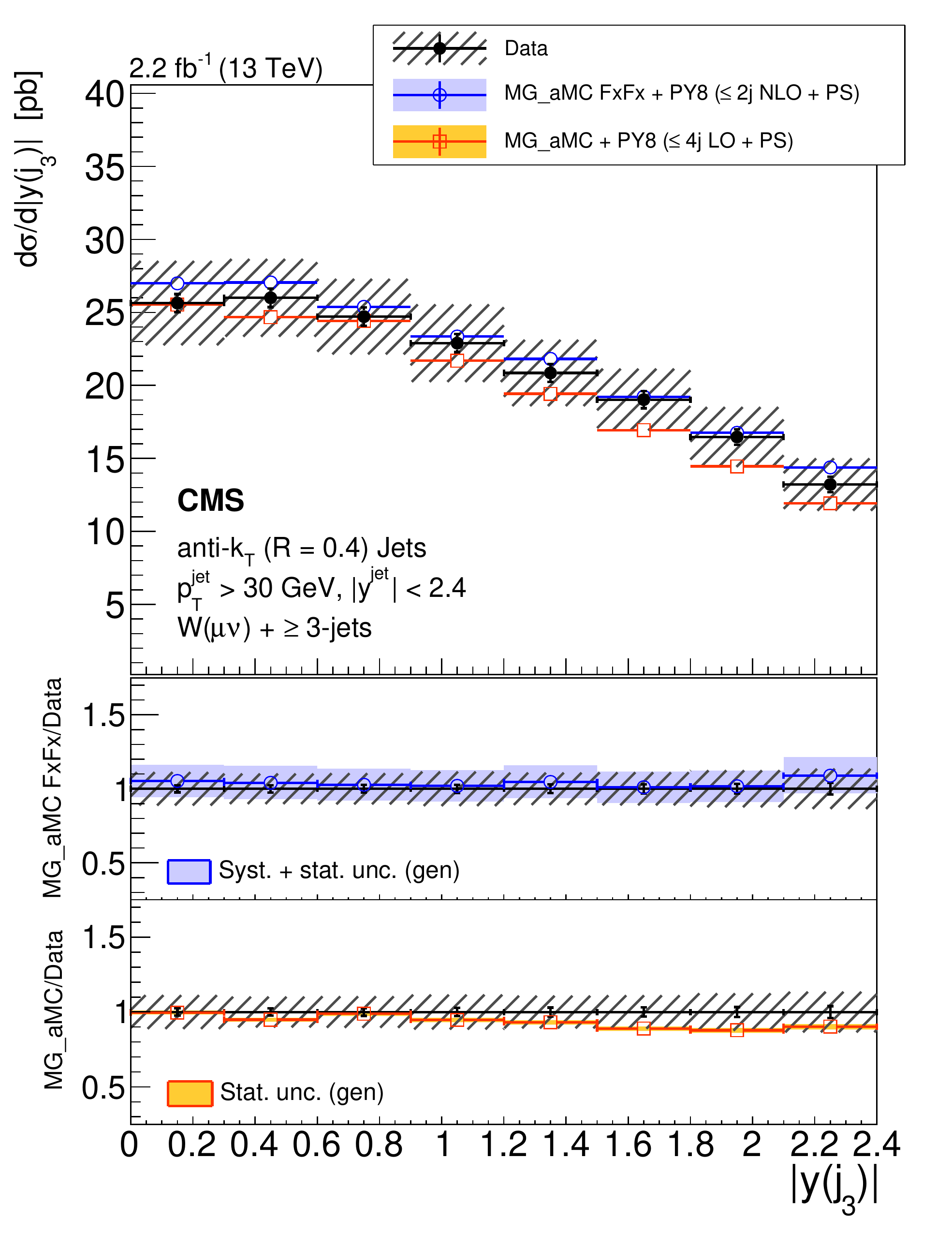}
\includegraphics[width=0.45\textwidth]{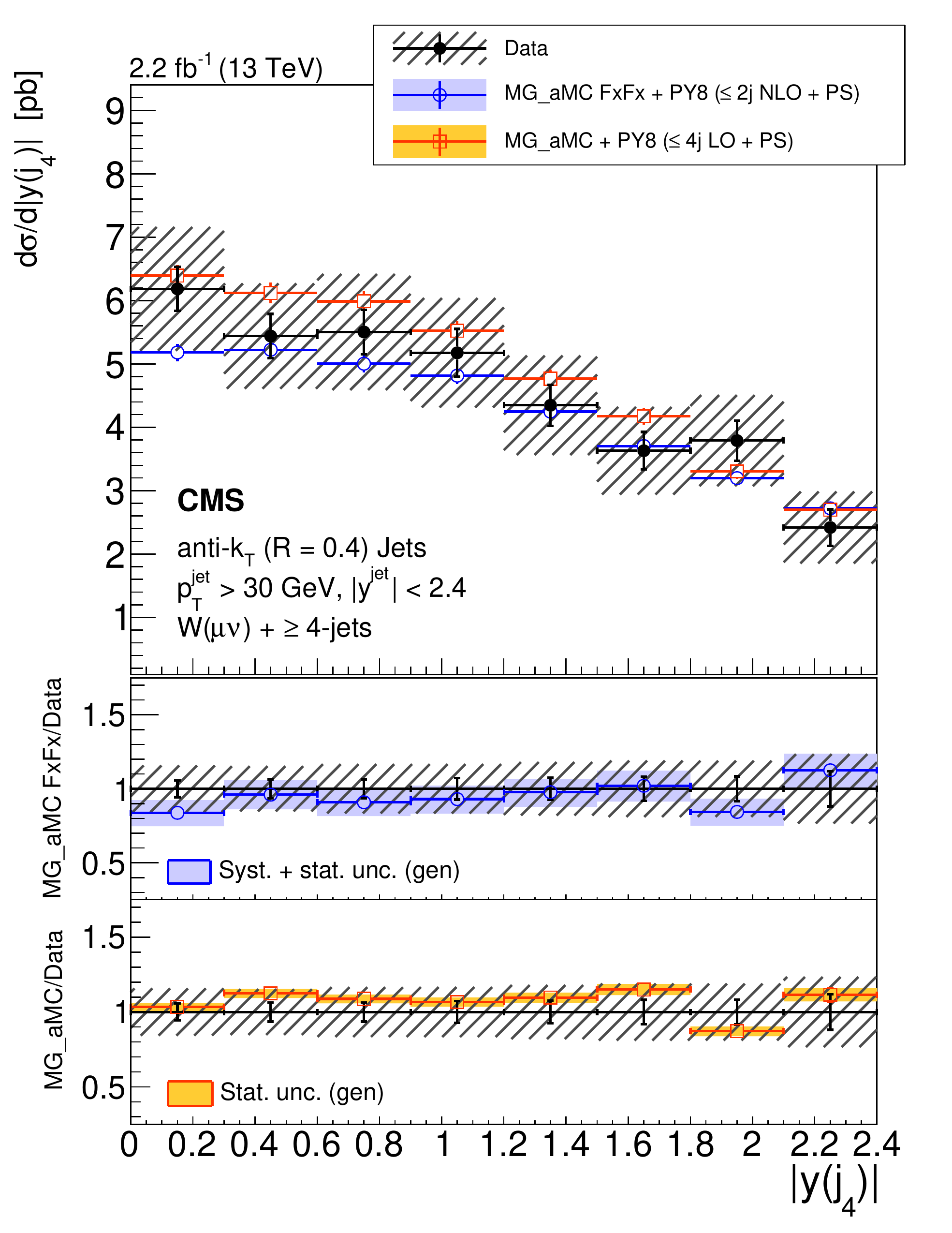}
\caption{Differential cross section measurement for the absolute rapidities of the four leading jets, shown from left to right for at least 1 and 2 jets (upper) and for at least 3 and 4 jets (lower) on the figures, compared to the predictions of NLO MG\_aMC FxFx and LO MG\_aMC. The NNLO prediction for $\PW$+1-jet is included in the first leading jet $|y|$. The black circular markers with the gray hatched band represent the unfolded data measurement and the total experimental uncertainty. The LO MG\_aMC prediction is given only with its statistical uncertainty. The bands around the NLO MG\_aMC FxFx and NNLO predictions represent their theoretical uncertainties including both statistical and systematic components. The lower panels show the ratios of the prediction to the unfolded data.}
\label{xsec_JetRapidity1to3_nlo}
\end{center}
\end{figure*}

The measured cross sections for inclusive jet multiplicities of 1--4 are compared with the predictions as a function of the jet $\pt$ ($|y|$) in Fig.~\ref{xsec_JetPt1to3_nlo} (Fig.\ref{xsec_JetRapidity1to3_nlo}). The measured cross sections as functions of the jet $\pt$ and $|y|$ are better described by the NLO MG\_aMC FxFx  prediction for all inclusive jet multiplicities and by the NNLO calculation for at least one jet. The LO MG\_aMC prediction exhibits a slightly lower trend in estimating data in contrast to NLO MG\_aMC FxFx and NNLO on jet $\pt$ and $|y|$ distributions, particularly at low $\pt$ and for inclusive jet multiplicities of 1--3.

\begin{figure*}[!htbp]
\begin{center}
\includegraphics[width=0.45\textwidth]{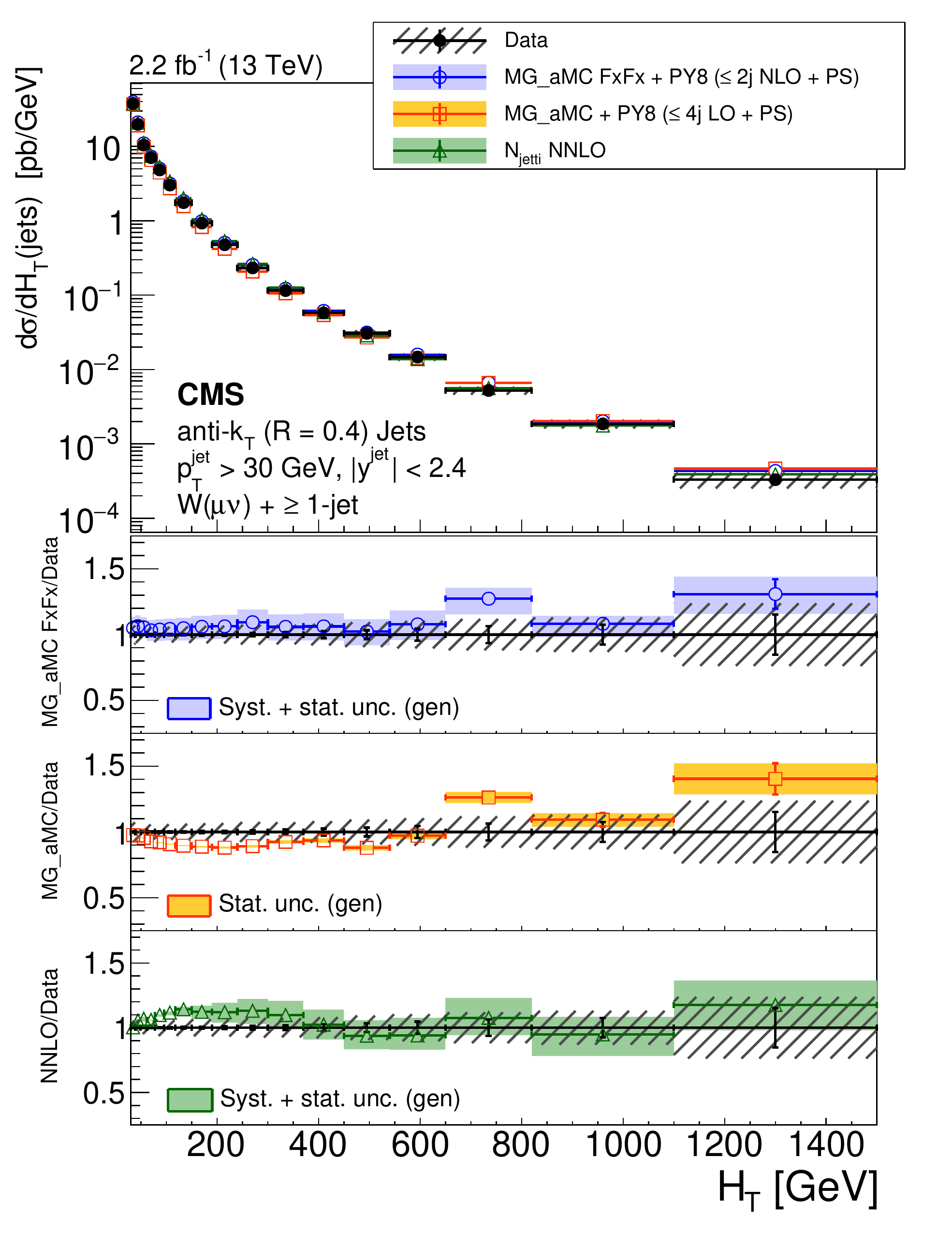}
\includegraphics[width=0.45\textwidth]{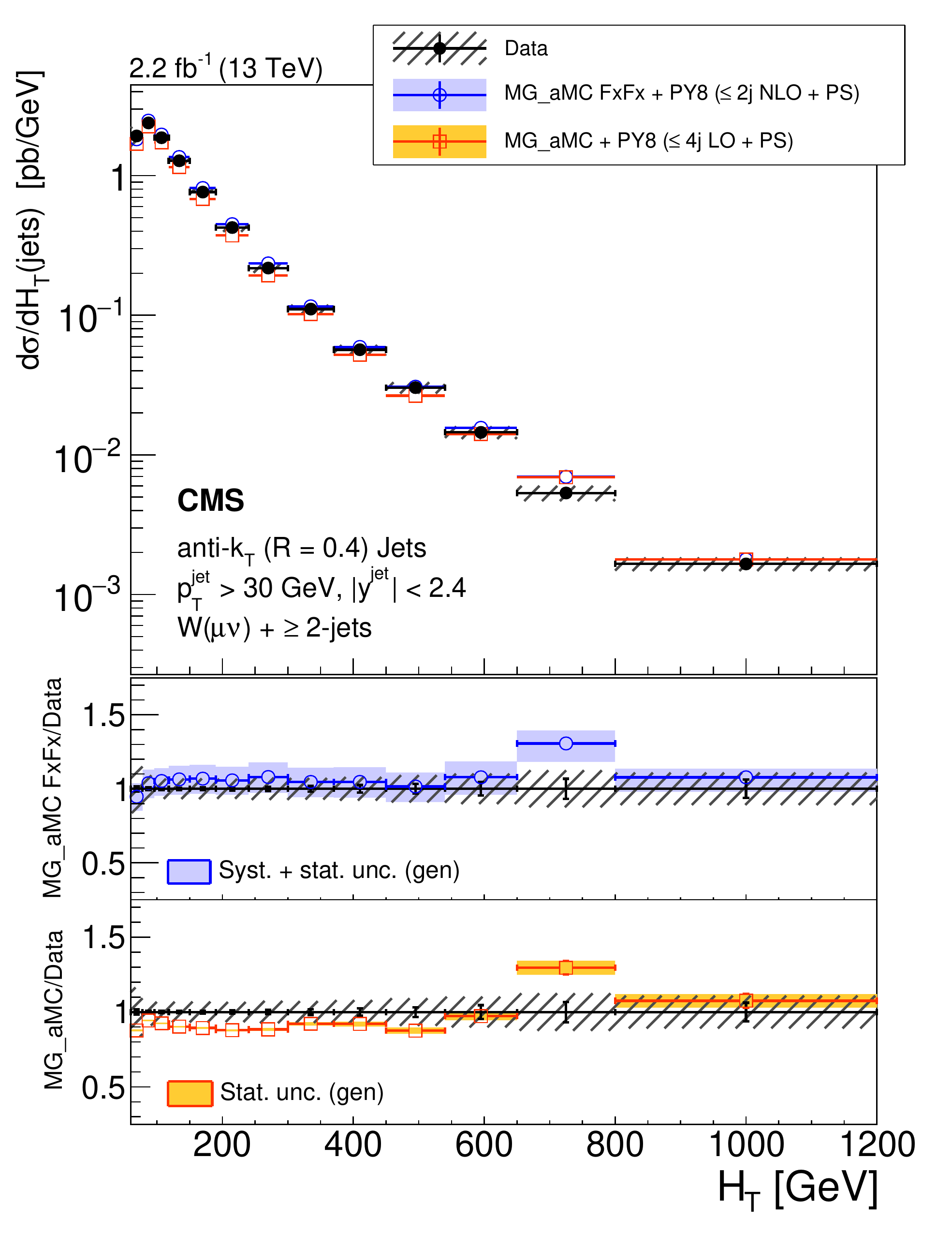}\\
\includegraphics[width=0.45\textwidth]{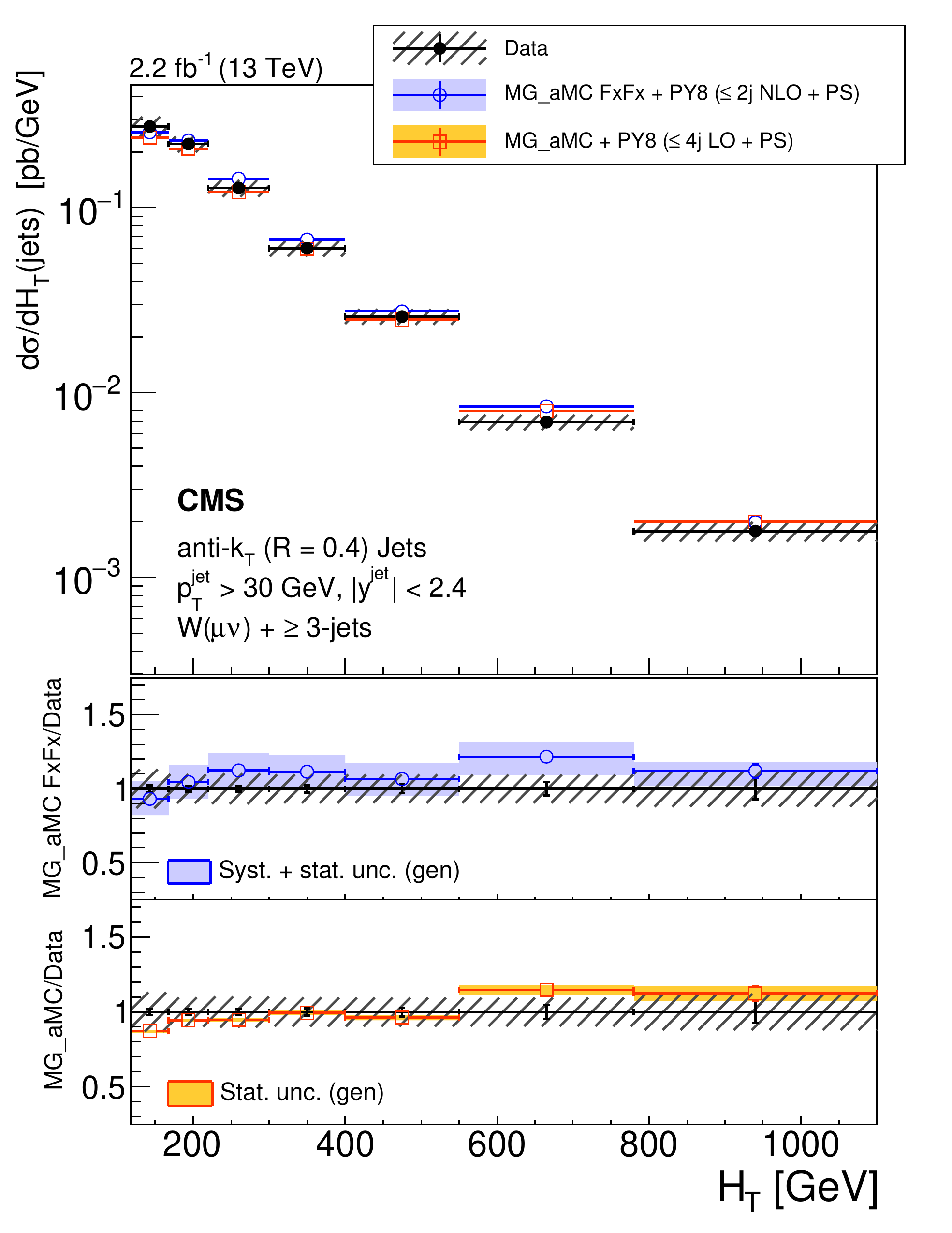}
\includegraphics[width=0.45\textwidth]{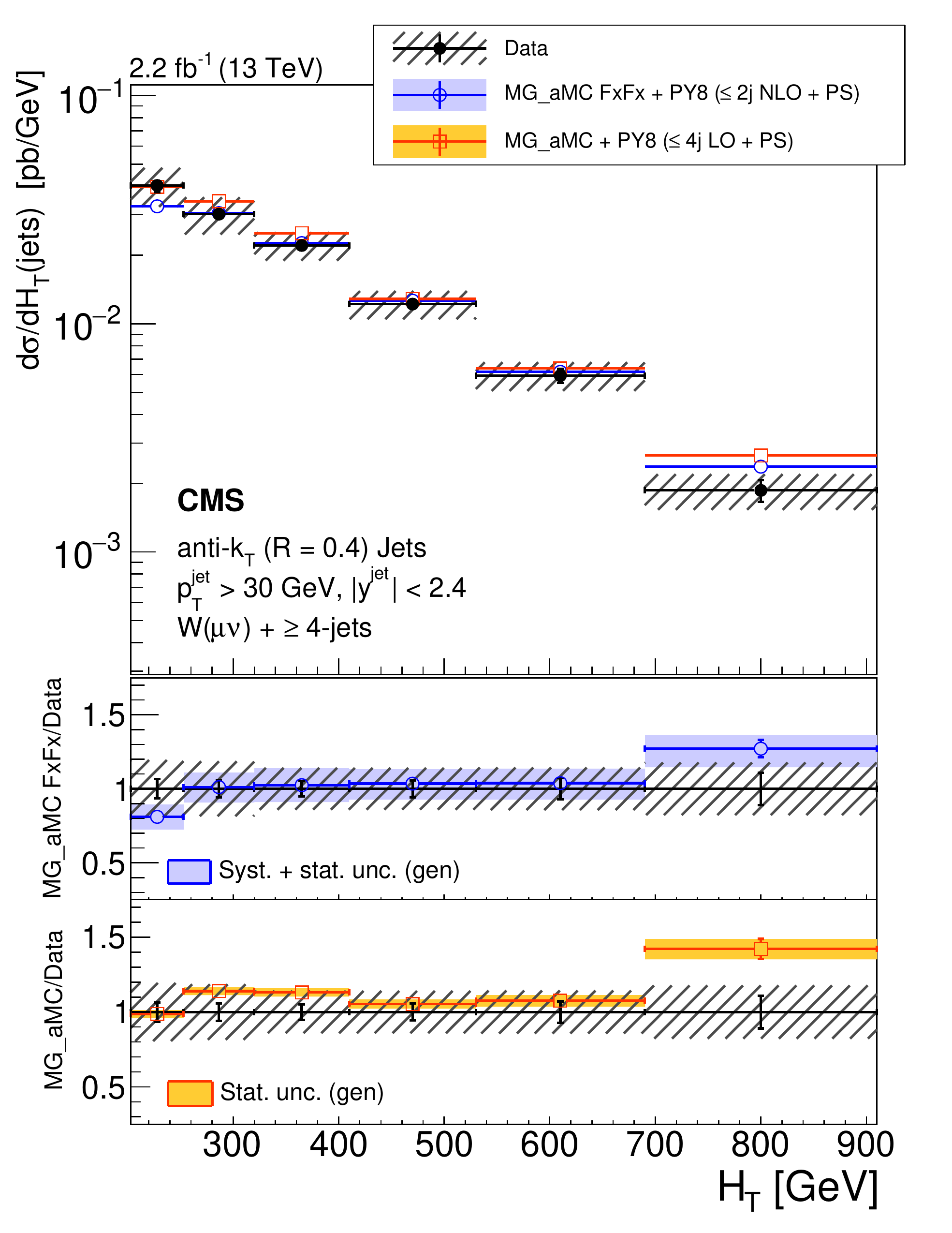}
\caption{Differential cross section measurement for the jets $\HT$, shown from left to right for at least 1 and 2 jets (upper) and for at least 3 and 4 jets (lower) on the figures, compared to the predictions of NLO MG\_aMC FxFx and LO MG\_aMC. The NNLO prediction for $\PW$+1-jet is included in the jets $\HT$ for one jet inclusive production. The black circular markers with the gray hatched band represent the unfolded data measurement and the total experimental uncertainty. The LO MG\_aMC prediction is given only with its statistical uncertainty. The bands around the NLO MG\_aMC FxFx and NNLO predictions represent their theoretical uncertainties including both statistical and systematic components. The lower panels show the ratio of the prediction to the unfolded data.}
\label{xsec_JetHT1to3_nlo}
\end{center}
\end{figure*}

The measured cross sections as functions of the $\HT$ variable of the jets, which is sensitive to the effects of higher order corrections, are compared with the predictions. The $\HT$ distributions for inclusive jet multiplicities of 1--4 are shown in Fig.~\ref{xsec_JetHT1to3_nlo}. The predictions are in good agreement with data for the $\HT$ spectra of the jets for all inclusive jet multiplicities, with the exception of LO MG\_aMC, which slightly underestimates the data at low $\HT$.

\begin{figure*}[!htbp]
\begin{center}
\includegraphics[width=0.45\textwidth]{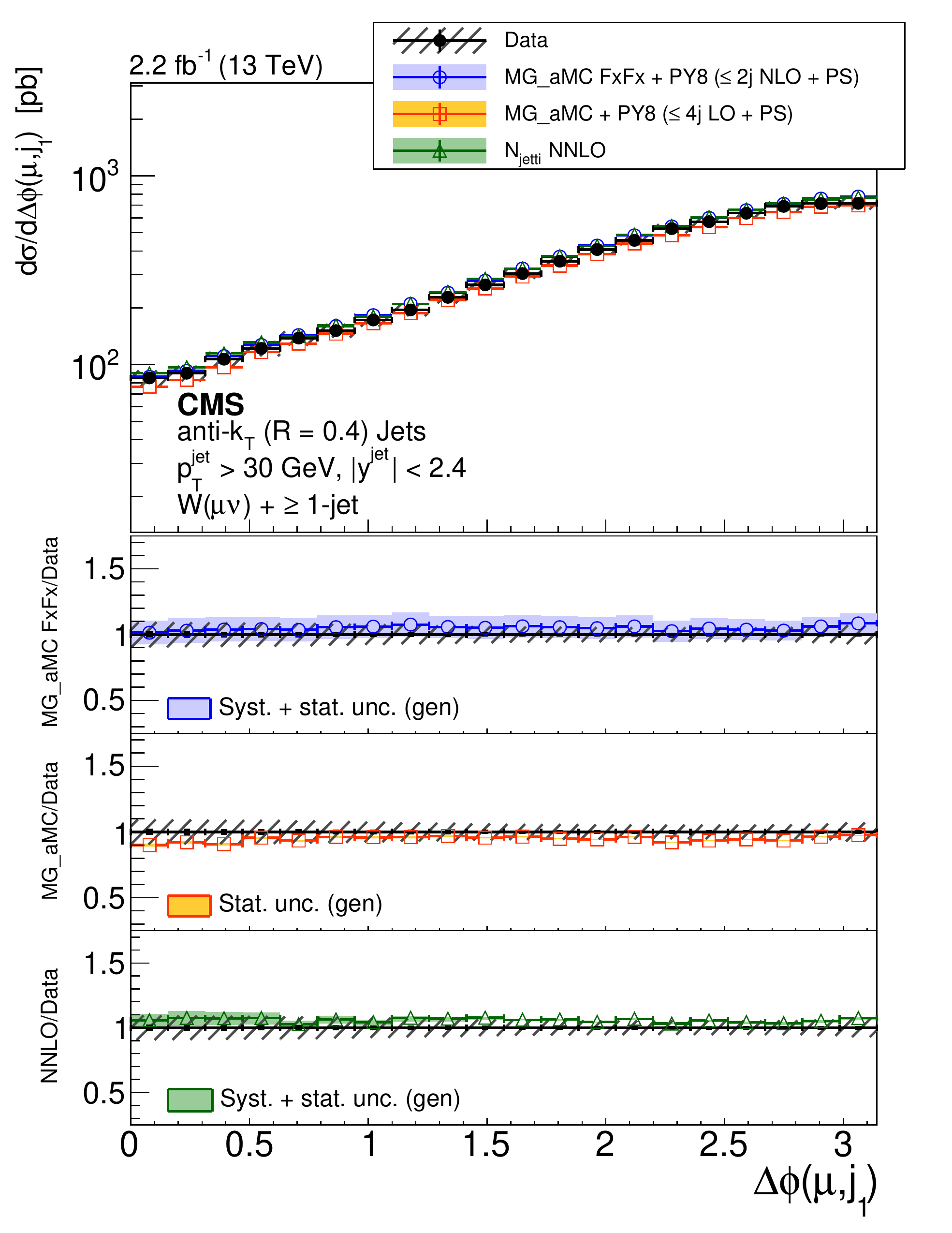}
\includegraphics[width=0.45\textwidth]{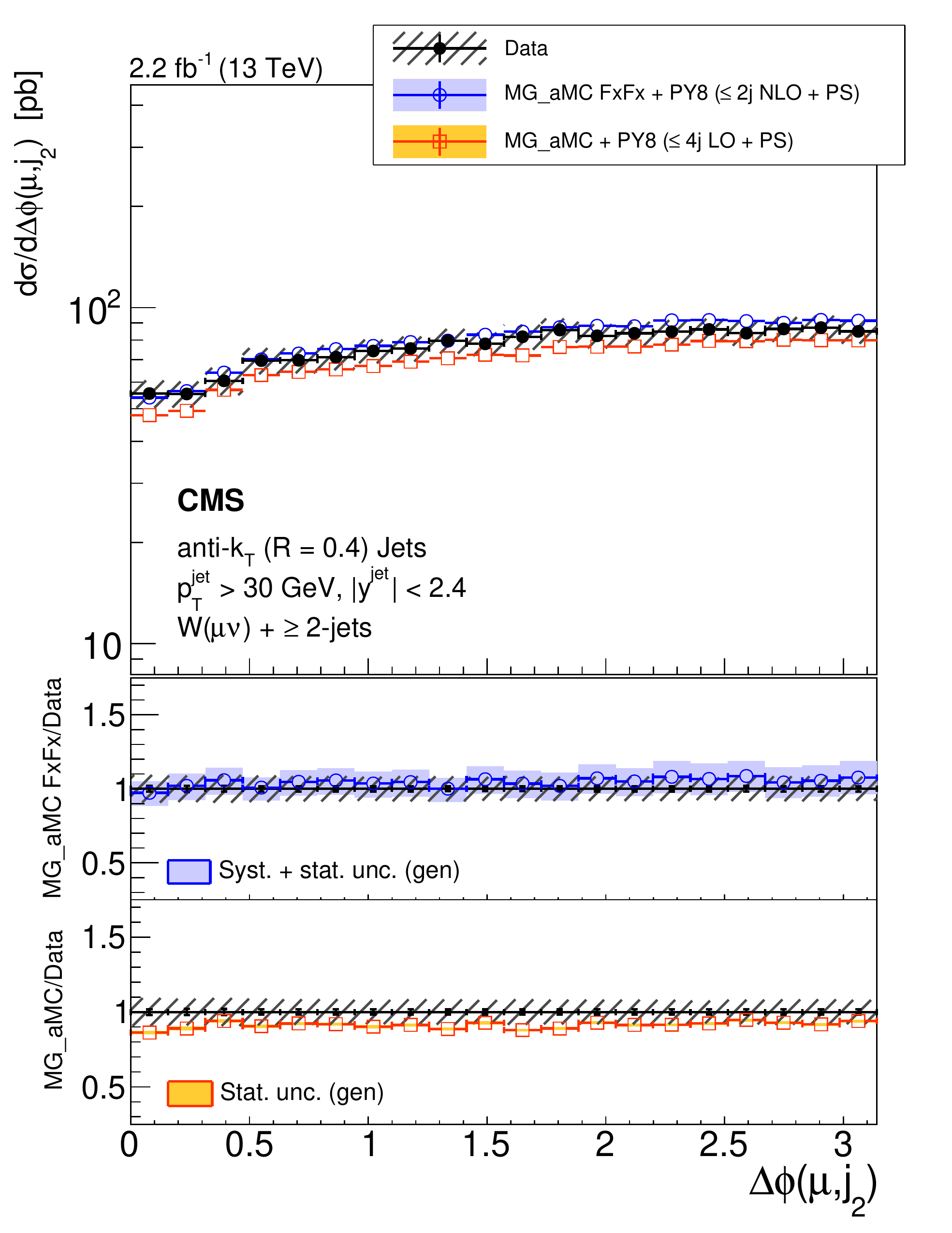}\\
\includegraphics[width=0.45\textwidth]{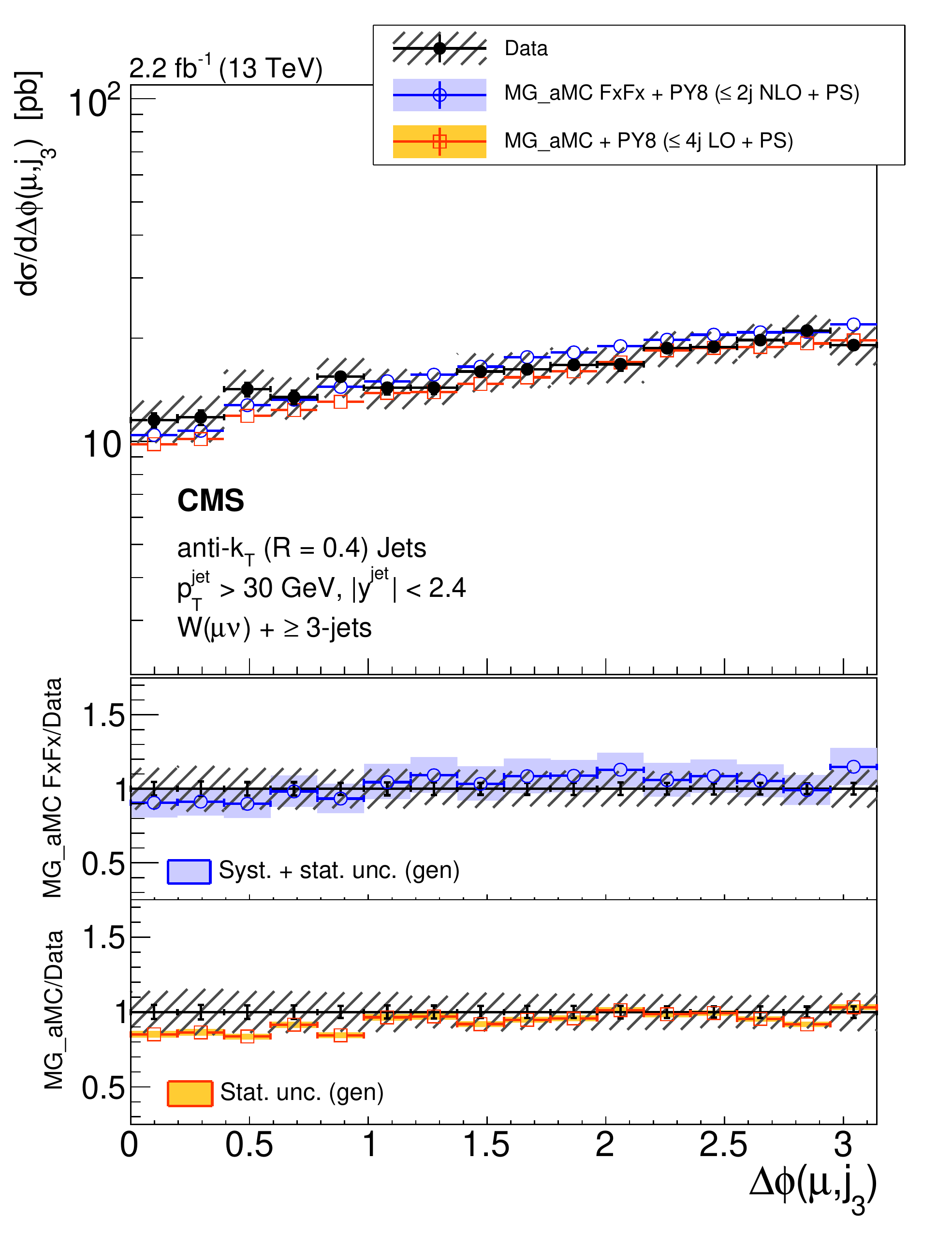}
\includegraphics[width=0.45\textwidth]{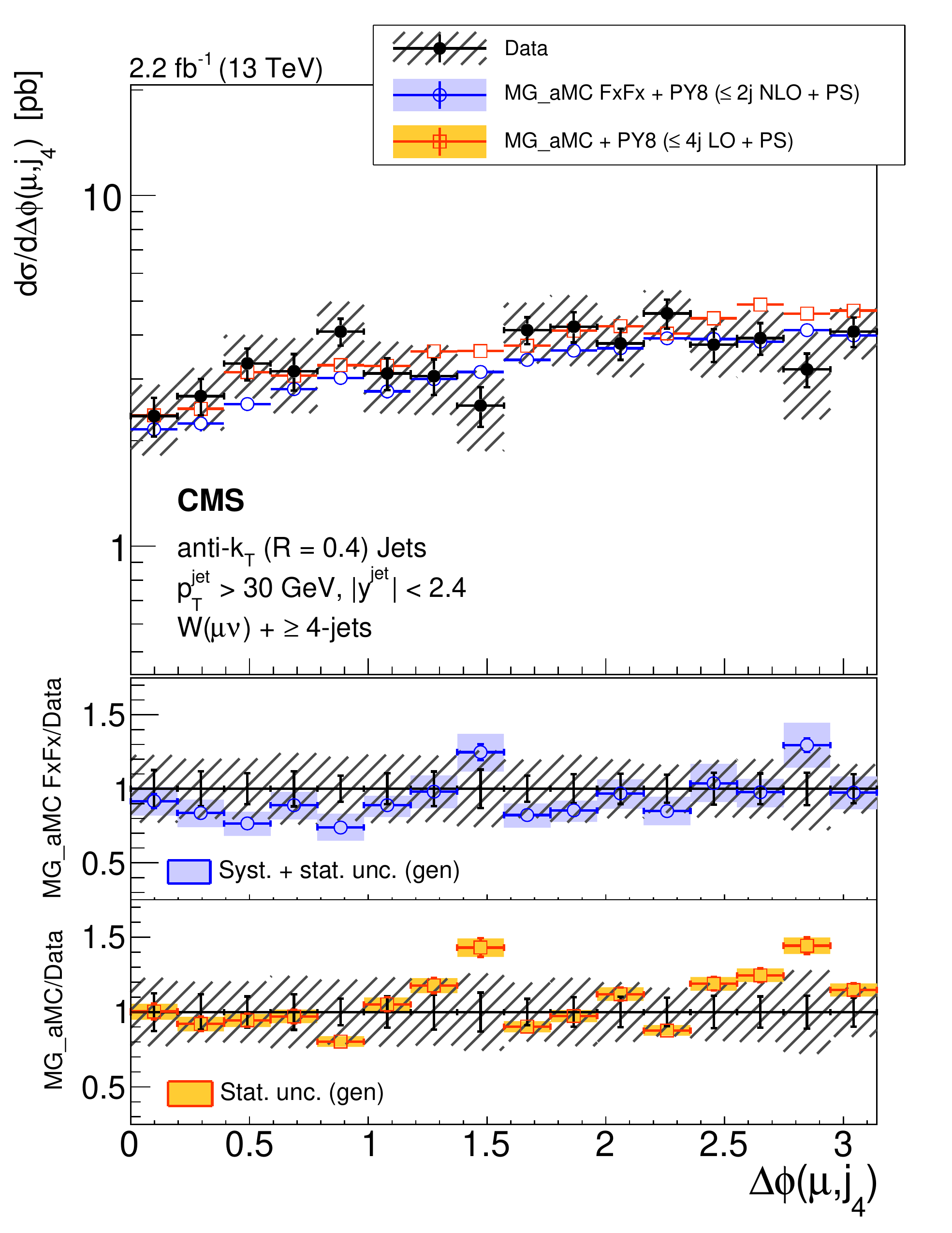}
\caption{Differential cross section measurement for $\Delta\phi(\text{$\PGm$, $j_{i}$})$, shown from left to right for at least 1 and 2 jets (upper) and for at least 3 and 4 jets (lower) on the figures, compared to the predictions of NLO MG\_aMC FxFx and LO MG\_aMC. The NNLO prediction for $\PW$+1-jet is included in $\Delta\phi(\text{$\PGm$, $j_{1}$})$ for one jet inclusive production. The black circular markers with the gray hatched band represent the unfolded data measurement and the total experimental uncertainty. The LO MG\_aMC prediction is given only with its statistical uncertainty. The bands around the NLO MG\_aMC FxFx and NNLO predictions represent their theoretical uncertainties including both statistical and systematic components. The lower panels show the ratio of the prediction to the unfolded data.}
\label{xsec_dphi1to4_nlo}
\end{center}
\end{figure*}

The differential cross sections are also measured as functions of angular variables: the azimuthal separation $\Delta\phi(\text{$\PGm$, $j_{i}$})$ between the muon and the jet for inclusive jet multiplicities of 1--4, and the angular distance between the muon and the closest jet $\Delta R(\text{$\PGm$, closest jet})$ in events with one or more jets. The measured $\Delta\phi(\text{$\PGm$, $j_{i}$})$ distributions are compared with the predictions in Fig.~\ref{xsec_dphi1to4_nlo} and they are well described within uncertainties. This observable is sensitive to the implementation of particle emissions and other nonperturbative effects modeled by parton showering algorithms in MC generators.

\begin{figure}[!htbp]
\begin{center}
\includegraphics[width=0.45\textwidth]{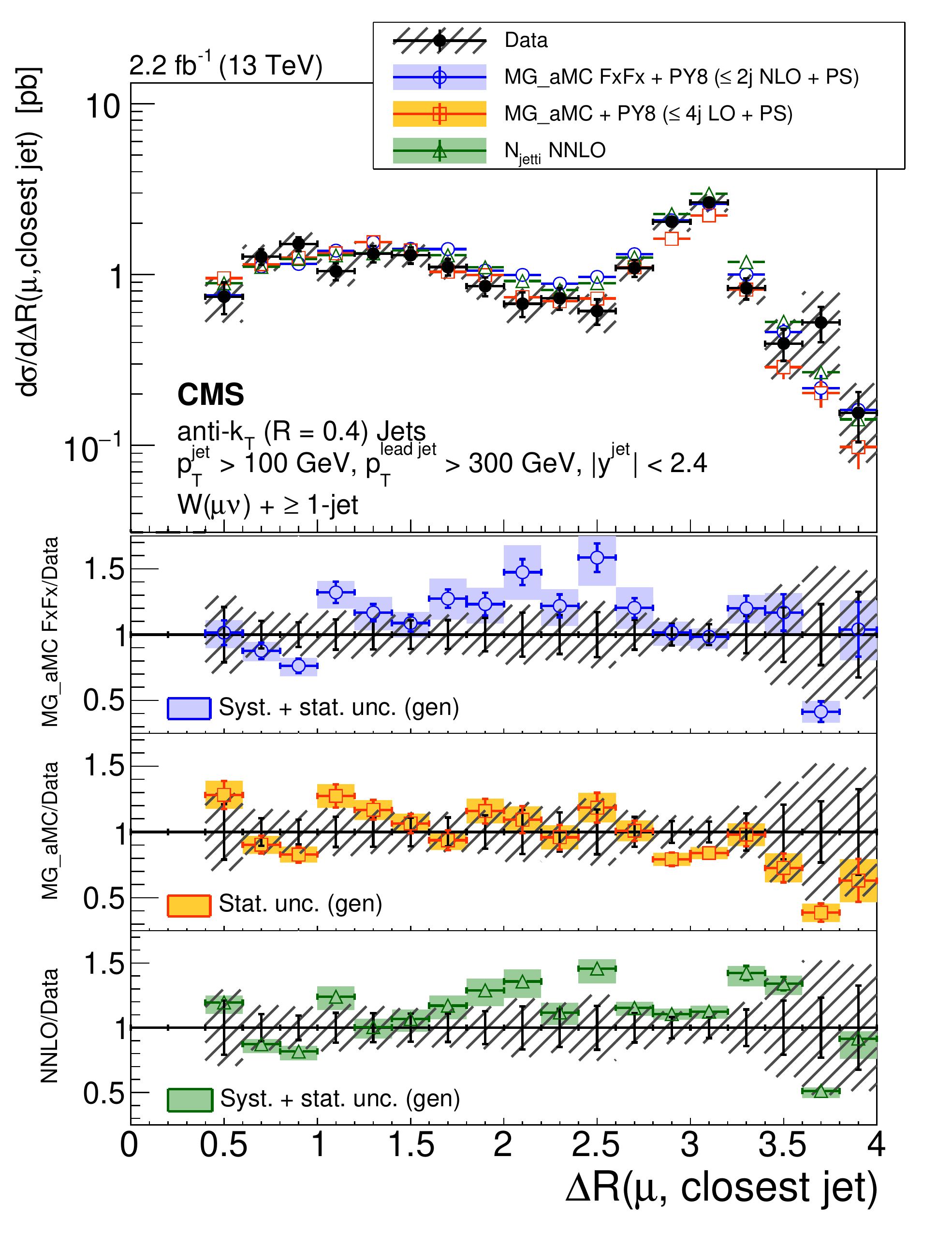}
\end{center}
\caption{Differential cross section measurement for $\Delta R(\text{$\PGm$, closest jet})$ for one jet inclusive production, compared to the predictions of NLO MG\_aMC FxFx, LO MG\_aMC, and the NNLO calculation. All jets in the events are required to have $\pt > 100\GeV $, with the leading jet $\pt > 300\GeV$. The black circular markers with the gray hatched band represent the unfolded data measurement and the total experimental uncertainty. The LO MG\_aMC prediction is given only with its statistical uncertainty. The bands around the NLO MG\_aMC FxFx and NNLO predictions represent their theoretical uncertainties including both statistical and systematic components. The lower panels show the ratio of the prediction to the unfolded data.}
\label{xsec_dr_nlo}
\end{figure}

The comparison of the measured $\Delta R(\text{$\PGm$, closest jet})$ with the predictions is shown in Fig.~\ref{xsec_dr_nlo}. This observable probes the angular correlation between the muon emitted in the $\PW$ boson decay and the direction of the closest jet. In the collinear region (small $\Delta R$ values), it is sensitive to the modeling of $\PW$ boson radiative emission from initial- or final-state quarks. The predictions are observed to be in fairly good agreement with data within the uncertainties, but there are some differences. Around $\Delta R =$ 2.0--2.5, in the transition between the region dominated by back-to-back $\PW$+$N$$\geq$$\text{1-jet}$ processes (high $\Delta R$) and the region where the radiative $\PW$ boson emission should be enhanced (low $\Delta R$), the NLO MG\_aMC FxFx prediction overestimates the measured cross section. In the high-$\Delta R$ region, the LO MG\_aMC prediction underestimates the data, which is consistent with the other observables.

\section{Summary}
\label{summary}

The first measurement of the differential cross sections for a $\PW$ boson produced in association with jets in proton-proton collisions at a center-of-mass energy of 13\TeV was presented. The collision data correspond to an integrated luminosity of 2.2\fbinv and were collected with the CMS detector during 2015 at the LHC.

The differential cross sections are measured using the muon decay mode of the $\PW$ boson as functions of the exclusive and inclusive jet multiplicities up to a multiplicity of six, the jet transverse momentum $\pt$ and absolute value of rapidity $|y|$ for the four leading jets, and the scalar $\pt$ sum of the jets $\HT$ for an inclusive jet multiplicity up to four. The differential cross sections are also measured as a function of the azimuthal separation between the muon direction from the $\PW$ boson decay and the direction of the leading jet for up to four inclusive jets, and of the angular distance between the muon and the closest jet in events with at least one jet.

The background-subtracted data distributions are corrected for all detector effects by means of regularized unfolding and compared with the predictions of \MADGRAPH{}5\_a\MCATNLO at leading-order (LO) accuracy (LO MG\_aMC) and at next-to-LO (NLO) accuracy (NLO MG\_aMC FxFx). The measured data are also compared with a calculation based on the $N$-jettiness subtraction scheme at next-to-NLO (NNLO) accuracy for $\PW$+1-jet production.

The predictions describe the data well within uncertainties as functions of the exclusive and inclusive jet multiplicities and are in good agreement with data for the jet $\pt$ spectra, with the exception of the LO MG\_aMC prediction, which underestimates the data at low to moderate jet $\pt$. The measured $\HT$ distributions are well modeled both by the NLO MG\_aMC FxFx prediction for all inclusive jet multiplicities and the NNLO calculation for $\PW$+1-jet. The LO MG\_aMC prediction underestimates the measured cross sections at low $\HT$. All predictions accurately describe the jet $|y|$ distributions and the cross sections as a function of the azimuthal correlation between the muon and the leading jet. The measured cross section as a function of the angular distance between the muon and the closest jet, which is sensitive to electroweak emission of $\PW$ bosons, is best described by the NNLO calculation.

\begin{acknowledgments}

We extend our thanks to Radja Boughezal for providing the NNLO predictions for this analysis~\cite{Boughezal:2015dva,Boughezal:2016dtm}.

We congratulate our colleagues in the CERN accelerator departments for the excellent performance of the LHC and thank the technical and administrative staffs at CERN and at other CMS institutes for their contributions to the success of the CMS effort. In addition, we gratefully acknowledge the computing centres and personnel of the Worldwide LHC Computing Grid for delivering so effectively the computing infrastructure essential to our analyses. Finally, we acknowledge the enduring support for the construction and operation of the LHC and the CMS detector provided by the following funding agencies: BMWFW and FWF (Austria); FNRS and FWO (Belgium); CNPq, CAPES, FAPERJ, and FAPESP (Brazil); MES (Bulgaria); CERN; CAS, MoST, and NSFC (China); COLCIENCIAS (Colombia); MSES and CSF (Croatia); RPF (Cyprus); SENESCYT (Ecuador); MoER, ERC IUT, and ERDF (Estonia); Academy of Finland, MEC, and HIP (Finland); CEA and CNRS/IN2P3 (France); BMBF, DFG, and HGF (Germany); GSRT (Greece); OTKA and NIH (Hungary); DAE and DST (India); IPM (Iran); SFI (Ireland); INFN (Italy); MSIP and NRF (Republic of Korea); LAS (Lithuania); MOE and UM (Malaysia); BUAP, CINVESTAV, CONACYT, LNS, SEP, and UASLP-FAI (Mexico); MBIE (New Zealand); PAEC (Pakistan); MSHE and NSC (Poland); FCT (Portugal); JINR (Dubna); MON, RosAtom, RAS, RFBR and RAEP (Russia); MESTD (Serbia); SEIDI, CPAN, PCTI and FEDER (Spain); Swiss Funding Agencies (Switzerland); MST (Taipei); ThEPCenter, IPST, STAR, and NSTDA (Thailand); TUBITAK and TAEK (Turkey); NASU and SFFR (Ukraine); STFC (United Kingdom); DOE and NSF (USA).

\hyphenation{Rachada-pisek} Individuals have received support from the Marie-Curie programme and the European Research Council and Horizon 2020 Grant, contract No. 675440 (European Union); the Leventis Foundation; the A. P. Sloan Foundation; the Alexander von Humboldt Foundation; the Belgian Federal Science Policy Office; the Fonds pour la Formation \`a la Recherche dans l'Industrie et dans l'Agriculture (FRIA-Belgium); the Agentschap voor Innovatie door Wetenschap en Technologie (IWT-Belgium); the Ministry of Education, Youth and Sports (MEYS) of the Czech Republic; the Council of Science and Industrial Research, India; the HOMING PLUS programme of the Foundation for Polish Science, cofinanced from European Union, Regional Development Fund, the Mobility Plus programme of the Ministry of Science and Higher Education, the National Science Center (Poland), contracts Harmonia 2014/14/M/ST2/00428, Opus 2014/13/B/ST2/02543, 2014/15/B/ST2/03998, and 2015/19/B/ST2/02861, Sonata-bis 2012/07/E/ST2/01406; the National Priorities Research Program by Qatar National Research Fund; the Programa Clar\'in-COFUND del Principado de Asturias; the Thalis and Aristeia programmes cofinanced by EU-ESF and the Greek NSRF; the Rachadapisek Sompot Fund for Postdoctoral Fellowship, Chulalongkorn University and the Chulalongkorn Academic into Its 2nd Century Project Advancement Project (Thailand); and the Welch Foundation, contract C-1845.

\end{acknowledgments}

\clearpage

\bibliography{auto_generated}

\cleardoublepage \appendix\section{The CMS Collaboration \label{app:collab}}\begin{sloppypar}\hyphenpenalty=5000\widowpenalty=500\clubpenalty=5000\input{SMP-16-005-authorlist.tex}\end{sloppypar}
\end{document}

%% file: SMP-16-005-authorlist.tex
\textbf{Yerevan Physics Institute,  Yerevan,  Armenia}\\*[0pt]
A.M.~Sirunyan, A.~Tumasyan
\vskip\cmsinstskip
\textbf{Institut f\"{u}r Hochenergiephysik,  Wien,  Austria}\\*[0pt]
W.~Adam, F.~Ambrogi, E.~Asilar, T.~Bergauer, J.~Brandstetter, E.~Brondolin, M.~Dragicevic, J.~Er\"{o}, M.~Flechl, M.~Friedl, R.~Fr\"{u}hwirth\cmsAuthorMark{1}, V.M.~Ghete, J.~Grossmann, J.~Hrubec, M.~Jeitler\cmsAuthorMark{1}, A.~K\"{o}nig, N.~Krammer, I.~Kr\"{a}tschmer, D.~Liko, T.~Madlener, I.~Mikulec, E.~Pree, D.~Rabady, N.~Rad, H.~Rohringer, J.~Schieck\cmsAuthorMark{1}, R.~Sch\"{o}fbeck, M.~Spanring, D.~Spitzbart, J.~Strauss, W.~Waltenberger, J.~Wittmann, C.-E.~Wulz\cmsAuthorMark{1}, M.~Zarucki
% ...
\vskip\cmsinstskip
% ...
1:~~Also at Vienna University of Technology, Vienna, Austria\\
% ...
